\documentclass[apj]{emulateapj}

\usepackage[utf8]{inputenc}
\usepackage{graphicx}
\usepackage{captcont}
\usepackage[english]{babel}
\usepackage{amssymb}
\usepackage{float}

\usepackage{color}

\slugcomment{{\sc Accepted to ApJ:} August 1, 2006} 

\usepackage[normalem]{ulem}

\slugcomment{}

\shorttitle{Terrestrial Planet Formation in Extra-Solar Systems}
\shortauthors{Izidoro et al.}

\begin{document}

\title{Terrestrial Planet Formation in the Presence of Migrating Super-Earths}

\author{André Izidoro\altaffilmark{1,2}}
\affil{University of Nice-Sophia Antipolis, CNRS, Observatoire de la Côte d’Azur, Laboratoire Lagrange, BP 4229,
06304 Nice Cedex 4, France.\\ Capes Foundation, Ministry of Education of Brazil, Brasília/DF 70040-020, Brazil.}
\email{izidoro.costa@gmail.com}

\author{Alessandro Morbidelli \altaffilmark{1}}
\affil{University of Nice-Sophia Antipolis, CNRS, Observatoire de la Côte d’Azur, Laboratoire Lagrange, BP 4229,
06304 Nice Cedex 4, France.}
\email{morbidelli@oca.eu}

\author{Sean.~N. Raymond\altaffilmark{3}}
\affil{CNRS and Université de Bordeaux, Laboratoire d'Astrophysique de Bordeaux, UMR 5804, F-33270, Floirac, France}
\email{rayray.sean@gmail.com}

\begin{abstract}

Super-Earths with orbital periods less than 100 days are extremely abundant around Sun-like stars. It is unlikely that these planets formed at their current locations.  Rather, they likely formed at large distances from the star and subsequently migrated inward.  Here we use N-body simulations to study the effect of super-Earths on the accretion of rocky planets.  In our simulations, one or more super-Earths migrates inward through a disk of planetary embryos and planetesimals embedded in a gaseous disk.  We tested a wide range of migration speeds and configurations.  Fast-migrating super-Earths ($\tau_{mig} \sim$0.01-0.1 Myr) only have a modest effect on the protoplanetary embryos and planetesimals. Sufficient material survives to form rocky, Earth-like planets on orbits exterior to the super-Earths'.  In contrast, slowly migrating super-Earths shepherd rocky material interior to their orbits and strongly deplete the terrestrial planet-forming zone.  In this situation any Earth-sized planets in the habitable zone are extremely volatile-rich and are therefore probably not Earth-like.

\end{abstract}

\keywords{Planets and satellites: formation; Methods: numerical}

\section{Introduction}

Close-in super-Earths are the most abundant known class of extrasolar planets.  Planets with sizes between 1 and 4 Earth radii with orbital periods smaller than 100 days are present around at least 30-50\% of Sun-like (GK spectral types) stars (Mayor et al. 2009, 2011; Howard et al. 2010, 2012) and perhaps close to 100\% (Chiang \& Laughlin 2013).  Many Earth-sized planets have been found.  Fressin et al. (2013) found that about 17\% of Sun-like stars have an Earth-sized planet with radii between 0.8 and 1.25 ${\rm R_{\oplus}}$ and orbital period smaller than 85 days. Petigura et al. (2013) found that $\sim$6\% of Sun-like stars have planets with radii between 1 and 2 ${\rm R_{\oplus}}$ and periods between 200-400 days (see also Dong \& Zhu 2013).  There are a few confirmed Earth-sized or slightly larger planets in the habitable zones of their host stars (e.g., Kepler-62 e -- Borucki et al 2013, and Kepler-186 f -- Quintana et al 2014).  

But size is not everything. Earth-sized planets around other stars -- even on Earth-like orbits -- might be significantly different from the Earth. It may be relevant to have a more restrictive definition of Earth-like planets.

The process of formation may be used to identify truly Earth-like planets. Raymond et al. (2014) defined two different classes of Earth-sized planets, called ``giant planetary embryos'' (or  ``mini-cores'') and terrestrial-like planets. Mini-cores are Earth-sized planets that formed rapidly when the gaseous disk was still around the star, and therefore they are likely to possess a primitive atmosphere. This makes them broadly different from the Earth. In contrast, extrasolar terrestrial-like planets are defined as planets that formed slowly, after the gaseous disk fully dissipated, through a phase of giant impacts among smaller planetary embryos,  in a process similar to that leading to the inner planets in our solar system (e.g. Chambers 2001).

Early papers proposing in situ formation (Hansen \& Murray, 2012, 2013; Chiang \& Laughlin, 2013) assumed that  the disk had a huge local density to start with. In situ formation can indeed reproduce many of the observed features of low-mass planets. However, the assumption of a extraordinarily high local surface density of the disk suffers from several inconsistencies.  First, it requires extremely massive disks very close to their stars (Raymond et al 2008, 2014; Chiang \& Laughlin 2013).  These disks must have a range of surface density slopes extending from extremely steep to extremely shallow slopes, a large fraction of which are incompatible with disk theory (Raymond \& Cossou 2014).  Second, in situ accretion inherently assumes that orbital migration is negligible.  However, the accretion timescale at such close-in orbital distances is so short that the planets must be fully formed long before the disappearance of the gaseous disk.  Assuming that no migration occurs essentially ignores 30 years of disk-planet studies that show the inevitability of orbital migration (e.g., Goldreich \& Tremaine 1980; Ward 1997).  Third, super-Earths that accrete in situ may have not been able to retain the H-He atmospheres that have been observed (Ikoma \& Hori, 2012). Finally, in situ accretion often fails at a conceptual level.  For instance, a ``minimum-mass'' disk that was generated from a given system often cannot actually lead to the formation of that system because other factors such as the width of feeding zones are not included in generating the disk (see Bolmont et al 2014 for the case of Kepler-186).
These problems have been at least partially alleviated by the proposal that the local disk density was enhanced due to pile-up of drifting material of various sizes, from grains to small-mass embryos (Hansen, 2014; Chatterjee, \& Tan; 2014; Boley \& Ford, 2013), down to some putative stopping radius.
If super-Earth planets in the vicinity of the star do form in this way, then the study that we propose in this paper obviously does not apply. However, we cannot fail noticing that this process did not happen at least in our solar system, whereas migration of Uranus and Neptune into the inner solar system would have occurred in absence of Jupiter and Saturn (Morbidelli, 2014).

Thus, in this paper we assume that most hot super-Earths formed via the inward migration model, namely we assume that they formed in the outer part of the disk and migrated close to the star due to the well-known planet-disk interactions (e.g. Terquem \& Papaloizou 2007; Cossou et al 2014; Raymond et al. 2014).  Although this model has not been fully tested (but see Cossou et al 2014), it offers a self-consistent picture that is consistent with other facets of planet formation (e.g., Papaloizou \& Terquem 2006; Armitage 2010), for instance: the fact that it may be easier to form big objects where water ice is available as solid material, that in our solar system the most massive bodies clearly formed in the outer disk, the low density of most characterized super-Earths (Marcy et al., 2014), the unavoidable character of planet migration. The migration model predicts that most final systems of super-Earths should be locked in mutual mean motion resonance (Rein, 2012; Rein et al., 2012; Ogihara \& Kobayashi, 2013; Ford, 2014; Zhang et al., 2014)). However, only a few observed systems may be in resonance.   Many multiplanet systems have orbital period ratios just slightly larger than first-order mean motion resonances (Lissauer et al., 2011; Fabrycky et al., 2014) and others are far from commensurability. This has been considered the main problem for the inward migration model.  However, several ideas have been proposed to explain this slightly off-resonance spacing between planet pairs. Exact period commensurabilities between these planets could be broken down as a result of dissipative effects and resonance repulsion in response to tidal damping (Lithwick \& Wu, 2012; Batygin \& Morbidelli, 2013; Delisle \& Laskar, 2014; but see also Petrovich et al., 2013), late instabilities in resonant chains triggered by the dispersal of the gaseous disk (Terquem \& Papaloizou 2007; Cossou et al 2014), planet interaction with a residual population of planetesimals (Chatterjee \& Ford, 2014) or late gravitational interaction between planets and their parent protoplanetary disk (Baruteau \& Papaloizou, 2013). Thus, the lack of exact resonances should not be taken as  evidence for in situ planet formation (Goldreich \& Schlichting, 2014).

Unfortunately, a true discrimination between the in situ and migration models is not currently possible.  This would require the knowledge of the chemical composition of the close-in super-Earths (see Table 1 of Raymond et al 2008). Planets with a large fraction of their mass in water should have formed beyond the snow-line and  suffered radial migration. If it contains a large amount of water, a planet will have a larger radius for a given mass (or lower mass for a given radius) than a dry planet (e.g., Fortney et al 2007).  We expect water-poor bodies to be mainly formed in situ (Raymond et al., 2008; Hansen \& Murray, 2012). Although the bulk densities of many low-mass planets have been measured (Marcy et al., 2014), a statistical discrimination between models requires higher precision than is currently feasible (see Selsis et al 2007).  In addition, as we show below, inward-migrating icy super-Earths can in some cases trigger the formation of close-in rocky planets, which makes the rocky versus icy nature of these objects a weaker discriminant of formation scenarios than hoped for. Also, the final mass-size relation can depend on what fraction of solid compared to gas the planets can accrete before reaching isolation (Chatterjee \& Tan, 2014).

In this paper we study the effect of migrating super-Earths on the formation of classical terrestrial planets.  We make two key assumptions.  First, we assume that hot super-Earths form by inward migration in the outer disk and reach their close-in orbits by inward migration.  For the reasons discussed above we think this represents a far more likely scenario than in situ accretion.  Second, we assume that hot super-Earths form first and therefore their migration has an influence on the accretion of rocky Earth-like planets. This second assumption is based on the expectation that rocky planets grow more slowly than icy/gaseous ones formed beyond the snow line because the latter could accrete from a radially wide disk of icy pebbles spiraling toward the ice line, while the former could only accrete from silicate grains and rocky planetesimals (Lambrechts et al., 2014; Kretke et al., 2014 communication at IAU symposium 310). Likewise, it is thought that Jupiter and Saturn (and even Uranus and Neptune) were fully-formed long before the terrestrial planets in the solar system and played an important role for Earth's formation (e.g., Morbidelli et al., 2012).

Previous studies have simulated the effect of migrating giant planets on terrestrial planet formation (Mandell \& Sigurdsson 2003; Fogg \& Nelson, 2005, 2007a, 2007b, 2009; Raymond et al., 2006; Lufkin et al 2006; Mandell et al, 2007). However, hot Jupiters are far less common than close-in super-Earths,  existing around only 0.5-1\% of Sun-like stars (Cumming et al 2008; Mayor et al., 2011; Howard et al., 2012; Batalha et al., 2013; Fressin et al., 2013). Thus, the formation of terrestrial-like planets in the context of migrating super-Earths is more interesting not only from an astrophysical perspective but also from an astrobiological point of view.

\subsection{Migrating Planets}

We now briefly review our current understanding of planet migration (see detailed reviews by Papaloizou \& Terquem 2006; Kley \& Nelson, 2012).  This will serve to motivate the setup of our simulations. 

There exist multiple different regimes of gas-driven planet migration.  The migration regime strongly depends on the mass of the migrating planets and the gas-disk properties (eg. Papaloizou et al., 2007). Jupiter mass planets open a gap and undergo ``type II'' migration which drives them to the vicinity of the star on a timescale comparable to that of the viscous evolution of the gaseous disk (Lin \& Papaloizou, 1986; Nelson et al., 2000). On the other hand, lower-mass planets, as super-Earths, are not able to open a gap and migrate in a different fashion, denominated ``Type I'' regime (Papaloizou \& Lin, 1984; Crida et al., 2006). 

The type I migration regime is complex and sensitive to the disk's properties. In a locally isothermal gas disk with surface density similar to the minimum-mass solar nebula (Weidenschilling, 1977; Hayashi et al, 1981)  a 10${ \rm M_{\oplus}}$ planet at 5 AU migrates all the way to the star in $\sim$ 0.01-0.1 Myr (Ward, 1997). This fast radial drift was considered for many years to be a major problem for the core accretion model of giant planet formation (Pollack et al., 1996), because a forming core would fall into the star before having the chance to accrete  enough atmosphere and become a giant planet. However, Paardekooper \& Mellema (2006) showed that in, non-isothermal disks with radiative transfer, type-I migration is slowed down and can even be reversed. This is due to slight changes in the gas density in the regions leading and trailing the planet along the horseshoe streamlines which exert a positive torque named ``entropy-driven coorbital torque''. This torque can in some cases exceed the net type-I Lindblad negative torque, driving outward migration.

Subsequent studies (Baruteau \& Masset 2008; Paardekooper \& Papaloizou 2008; Kley \& Crida, 2008; Bitsch \& Kley, 2011) found that outward migration can only occur in the inner part of the disk.  The outer part of the disk behaves as an isothermal disk. Consequently, outward migration must stop at an orbital radius where the entropy-driven torque and the type I Lindblad torque cancel out (Paardekooper \& Mellema, 2008). This orbital radius, which is dependent on planet mass, is called a ``zero-torque radius''.  As the disk evolves, zero-torque regions shift and move inward (Lyra et al. 2010). Thus, upon reaching the zero-torque radius a planet will migrate slowly inward, as the disk loses mass by viscous accretion and photoevaporation. Following this process, Lyra et al., (2010) concluded that low-mass planets avoid the catastrophic infall of planetary cores predicted by isothermal models and are instead delivered within $\sim 1$ AU from the star at the time of the ultimate disappearance of the gas.
Super-Earths (and mini-Neptunes) are observed in the vicinity of at least 30-50\% of the stars (e.g., Howard et al 2010, 2012; Mayor et al 2011).  If these planets formed in situ via the pileup of migrating material (Hansen, 2014; Chatterjee, \& Tan; 2014; Boley \& Ford, 2013) one should expect that terrestrial planets with smaller masses should be present further out, possibly reaching the habitable zone. However, if super-Earths formed in the outer disk and migrated to their current location,   the effect of their migration must be taken into account when considering the formation of Earth-sized planets in the habitable zone.    
Bitsch et al., (2014) showed that type I migration in real disks may be even more complex. The  dependence of the net total torque on planetary mass and on the  disk's dust-gas ratio demonstrated in their study, indicated that the zero-torque radius might disappear at some stage during the disk's evolution. Consequently, a planet initially at the zero-torque radius might be let free to migrate toward the star at a fast type I like migration rate when the zero-torque radius disappears. If this is correct, the real evolution of small-mass planets is totally different from that described in Lyra et al., (2010).

To summarize, it is not currently possible to quantitatively and reliably model the migration history of super-Earths toward the central star.  In this study we explore a range of migration rates  for the super-Earths (fast and slow) and different timescales for the gas-disk dissipation. Planets migrating inward at different timescales may cross the terrestrial region at different ages of the gas disk, and also disturb the protoplanetary bodies in this region in different as a result of distinct migration speeds. We identify correlations between the different migration scenarios and the resulting state of the protoplanetary disk of planetary embryos and planetesimals potentially leading to the terrestrial planet formation in the habitable zone of the star.

\vskip 5 pt

The layout of the paper is as follows. In section 2 we describe our model and the initial conditions of our simulations. In Section 3 we outline our simulations. In Section 4 we present the results of the simulations. In Section 5 we discuss the composition and habitability of terrestrial exoplanets in the context of our simulations.  In Section 6 we summarize our conclusions.

\section{Methods}

We distinguish two different phases of planet formation: pre- and post-gas disk dissipation (Armitage, 2010). Gaseous protoplanetary disks around young stars are believed to last no more than 10 Myr (Haisch et al., 2001; see Soderblom et al., 2014 for a review). Over this timescale, gas effects play a very important role for the dynamics of building blocks of planets regulating, for example, the accretion rates of planetesimals onto protoplanets (e.g. Safronov, 1972). Gas-driven planet migration also occurs during this phase.

Our simulations are divided into two parts. In the first part we set up a disk of planetesimals and planetary embryos. We study how this disk evolves in response to the migration through the habitable region of one or more super-Earths. Various gas effects on planetesimals and planetary embryos are taken into account (specified below). The gas is removed exponentially during the first phase. The second part of the simulations evolves the population of surviving embryos and planetesimals in a gas-free environment for a few million years. The amount of material near the habitable zone gives an indication of whether terrestrial planet formation in that region may be possible or not.

Below, we describe the initial set up of the disk of embryos and planetesimals, of the disk of gas, and of the gas effects (drag, tidal damping and migration).

\subsection{Disk of Solids}

Our simulations start from a disk of planetary embryos and planetesimals predicted by runaway and oligarch models of accretion (Kokubo \& Ida, 1998, 2000, 2002). The central host star is assumed to have the same mass as our Sun.  The initial protoplanetary disk extends from 0.3 AU to ${\rm \sim 4.8 AU}$, and its mass is equally divided between planetesimals and planetary embryos. The disk's radial mass distribution follows a surface density profile given by 
\begin{equation}
{\rm \Sigma(r)_{d}=
\Sigma_{1_{d}} \left(\frac{r}{1 AU}\right)^{-3/2}.\hspace{.3cm}}
\end{equation}
where we have assumed a surface density at 1 AU, $\Sigma_{1d}= 13 gcm^{-2}$.  This corresponds to a protoplanetary disk  nearly twice as massive as the minimum-mass solar nebula (Weidenschilling 1977, Hayashi, 1981), with $\sim$10 Earth masses of solid material between 0.3 and ${\rm \sim 4.8 AU}$ AU. The individual embryo's mass scales as $r^{3(2-\alpha)/2}\Delta^{3/2}$ (Kokubo \& Ida 2002; Raymond et. al., 2005), where $\alpha$ is the surface density profile index and $\Delta$ is the number of mutual Hill radii. In our case $\alpha=$1.5 (Eq. 1) and $\Delta$ is assumed to be 5-10 (Kokubo \& Ida, 2000). Planetesimals' individual masses are set to be ${\rm 10^{-8}}$ solar masses  (= 0.0033 Earth masses). In total, there are about 1500 planetesimals and $\sim$80 planetary embryos. While planetary embryos are allowed to gravitationally interact with all bodies in the system, planetesimals do not feel each other but are allowed to interact with other objects (planetary embryos, central star and super-Earths).  The initial orbital inclinations of all bodies in the disk were chosen randomly from ${\rm 10^{-4}}$ to ${\rm 10^{-3}}$ degrees, eccentricities were set initially to zero and mean anomalies were taken randomly between $\rm 0^{\circ}$ and $\rm 360^{\circ}$. The arguments of periastron and longitudes of ascending nodes of all objects were initially set to zero. Figure 1 shows  one of our initial conditions. In this figure planetesimals are the filled circles with mass equal to 0.0033 Earth masses. Planetary embryos are represented by open circles with masses ranging from Moon-size to Mars-size. For each scenario we studied, we ran at least 3 simulations with slightly different initial conditions for the protoplanetary embryos and planetesimals. When generating these different initial conditions, we used different randomly generated values for $\Delta$ in the  5-10 range and also for the orbital inclinations and mean anomalies of planetary embryos and planetesimals.

\begin{figure}[h]
\centering
\includegraphics[scale=.7]{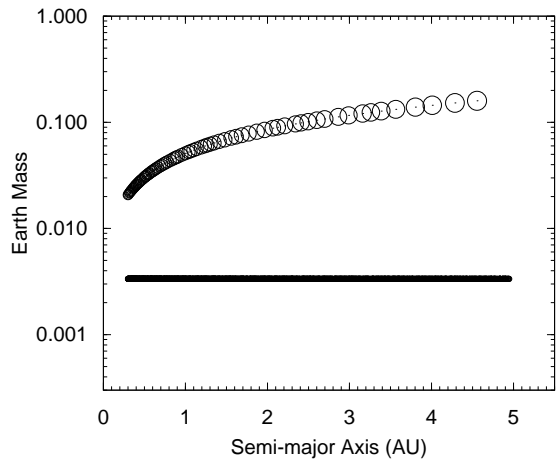}
\caption{Initial distribution of 82 planetary embryos  and 1500 planetesimals representative of our set of initial conditions. The masses of planetesimals are smaller than 0.004 Earth masses (${\rm \sim 0.0033 M_{\oplus} }$.)}
\end{figure}

\subsection{Disk of Gas}

We include a power-law disk model to represent the surface density of gas.  The gas surface  density is taken to be
\begin{equation}
{\rm \Sigma_{g}(r)=
\Sigma_{1g} \left(\frac{r}{1 AU}\right)^{-3/2},\hspace{.3cm}}
\end{equation}
where $r$ is the distance to the star  and ${\rm \Sigma_{1g}=3400gcm^{-2}}$ is the reference surface density at 1 AU. Thus as for the surface density in solids, the gas surface density corresponds to a model twice as massive as the minimum-mass solar nebula. The disk radial temperature is considered as in Hayashi (1981),
\begin{equation}
{\rm T(r) = 280 \left(\frac{r}{1AU}\right)^{-1/2} K}
\end{equation}
where r also represents the distance to the star and the temperature is considered vertically uniform in the disk. The corresponding isothermal sound velocity is given by
\begin{equation}
{\rm c_{s}(r) = 9.9\times 10^4 \left(\frac{r}{1AU}\right)^{-1/4} cm\hspace{0.2mm}s^{-1}}
\end{equation}
and the scale height is 
\begin{equation}
{\rm h(r) = h_1   \left(\frac{r}{1AU}\right)^{1.25}  AU}
\end{equation} 
where ${\rm h_1= 0.047}$ is the half-thickness of disk at 1 AU. The density in the midplane is given by

\begin{equation}
{\rm \rho_{g}(r) =  \frac{\Sigma(r)_{g}}{\sqrt{\pi}h}= 2.9\times 10^{-9}\left(\frac{r}{1AU}\right)^{-11/4}g cm^{-3}} 
\end{equation} 

To mimic the gas disk dissipation due to viscous evolution and photoevaporation, we assumed a simple exponential decay for the gas surface density, given by ${\rm Exp(-t/\tau_{gas})}$, where ${\rm t}$ is the time and ${\rm \tau_{gas}}$ is the gas dissipation timescale. Simulations were carried out considering values for ${\rm \tau_{gas}}$ between  1 and 10 Myr. For simplicity, in all simulations, the gas is assumed to dissipate instantaneously at  t$=\tau_{diss}$ and the dynamical system starts to evolve in a gas-free environment. We fix $\tau_{diss}$  to be the largest between $\tau_{gas}$ and the super-Earth migration timescale.

\subsubsection{Gas Effects} 

\paragraph{Aerodynamic gas drag on planetesimals}

Gas drag transfers angular momentum between the gas and the planetesimals.  When gas orbits at a sub-Keplerian speed, it causes planetesimals to spiral inward. The aerodynamic gas drag implemented in our simulations follows the formalism of Adachi et al., (1976). The deviation of the angular velocity of the gas from the Keplerian angular velocity is
\begin{equation}
\Omega_g = \Omega_k(1 - \eta)^{1/2}
\end{equation} 
where ${\rm \Omega_k}$ is the Keplerian orbital frequency and  $\eta$ is the ratio of the gas pressure gradient to the stellar gravity in the radial direction  which is 
\begin{equation}
{\rm \eta = \frac{11}{16}\left(\frac{h}{r}\right)^2.}
\end{equation}

The gas drag acceleration is given by 

\begin{equation}
{\rm {\bold a}_{drag} =- \frac{3C_d \rho_g v_{rel} \bold{v}_{rel}}{8 \rho_p R_p}}
\end{equation}
where ${\rm C_d}$ is the drag coefficient, ${\rm v_{rel}}$ is the relative velocity of the object with respect to the surrounding gas, and ${\rm \rho_p}$ and ${\rm R_p}$ are the planetesimal's bulk density and radius, respectively. The coefficient gas drag is implemented as in Brasser et al., (2007).  As observed in Eq. 9, the planetesimal radius is the key factor governing the magnitude of the aerodynamic gas drag. The nominal planetesimal radius assumed in our simulations was 100 km, however, for comparison, we also performed simulations considering planetesimals with radii equal to 10 and 1000 km.

\paragraph{Tidal interaction of protoplanetary embryos with the gas} 

Planetary embryos and planetesimals interact differently with the gaseous protoplanetary disk. While small planetesimals only feel the headwind of the gas, planetary embryos are massive enough to gravitationally excite density waves. Consequently, they feel torques exerted by those waves (Goldreich \& Tremaine 1980; Ward,1986; Tanaka et al., 2002; Tanaka \& Ward, 2004). For a planetary embryo, evolving in a slightly eccentric and/or inclined orbit these density waves can be divided into three groups. The first group acts even if the planet's inclination and eccentricities are zero. They are spiral density waves launched at the Lindblad resonances, which transfer angular momentum to the disk and force the embryo to migrate inward. The second and third groups are made of eccentricity and bending waves, which damp the embryo's eccentricity and inclination, respectively (Tanaka \& Ward, 2004).

To include the effects of type I migration, eccentricity and inclination damping on embryos, we followed the formalism of Tanaka et al., (2002) and Tanaka \& Ward (2004), modified  by Papaloizou and Larwood (2000), and Cresswell \& Nelson (2006; 2008) to include the case of large eccentricities. The timescales for semimajor axis, eccentricity and inclination damping are given  by ${\rm t_m}$, ${\rm t_e}$ and ${\rm t_i}$, respectively. Their values are:

\begin{equation}
\footnotesize
{\rm t_m = \frac{2}{2.7 + 1.1\alpha} \left(\frac{M_{\odot}}{m_p}\right)  \left(\frac{M_{\odot}}{\Sigma_g {a_p}^2}\right)\left(\frac{h}{r}\right)^2 \left(\frac{1+ \left(\frac{er}{1.3h}\right)^5}{1-\left(\frac{er}{1.1h}\right)^4}\right)\Omega_k^{-1}}, 
\end{equation}

\begin{equation}
\scriptsize
{\rm t_e = \frac{t_{wave}}{0.780} \left(1-0.14\left(\frac{e}{h/r}\right)^2 + 0.06\left(\frac{e}{h/r}\right)^3    + 0.18\left(\frac{e}{h/r}\right)\left(\frac{i}{h/r}\right)^2\right),}
\end{equation}
and
\begin{equation}
\scriptsize
{\rm t_i = \frac{t_{wave}}{0.544} \left(1-0.3\left(\frac{i}{h/r}\right)^2 + 0.24\left(\frac{i}{h/r}\right)^3    + 0.14\left(\frac{e}{h/r}\right)^2\left(\frac{i}{h/r}\right)\right)}
\end{equation}

where
\begin{equation}
{\rm t_{wave} = \left(\frac{M_{\odot}}{m_p}\right)  \left(\frac{M_{\odot}}{\Sigma_g a^2}\right)\left(\frac{h}{r}\right)^4 \Omega_k^{-1}}
\end{equation}
and ${\rm M_{\odot}}$, ${\rm a_p}$, ${\rm i}$, ${\rm e}$ are the solar mass and the embryo's semimajor axis, inclination and eccentricity, respectively. $\alpha$ is the gas disk surface density profile index, which in our case is equal to 1.5 (Eq. 2).

To model the damping of semimajor axis, eccentricity, and inclination, we implemented synthetic accelerations as defined in Cresswell \& Nelson (2008):
\begin{equation}
{\rm \bold{a}_m = -\frac{\bold{v}}{t_m}}
\end{equation}

\begin{equation}
{\rm \bold{a}_e = -2\frac{(\bold{v.r})\bold{r}}{r^2 t_e}}
\end{equation}

\begin{equation}
{\rm \bold{a}_i = -\frac{v_z}{t_i}\bold{k},}
\end{equation}
where ${\rm \bold{k}}$ is the unit vector in the z-direction.

\section{Numerical Simulations}

Our simulations were run with the Symba integrator (Duncan et al., 1998) adopting a 2-day time step. The code has been slightly changed to include aerodynamic gas drag, type I migration, and tidal damping of the embryos' eccentricities and inclinations as explained before. Collisions are considered to be inelastic, resulting in a merging event that conserves linear momentum. During the simulations, planetesimals and protoplanetary embryos that reach heliocentric distances smaller than 0.1 AU are  assumed to collide with the central body.  Planetesimals or protoplanetary embryos are removed from the system if ejected beyond 100 AU of the central star.

\subsection{The Different Models}

Gas drag -- both aerodynamic and tidal -- plays a very important role both during and after the migration of a large planet (Fogg \& Nelson, 2005; 2007a; Raymond et al., 2006; Mandell et al 2007). Different approaches have been used when considering the gas dissipative effects on planetesimals and protoplanetary embryos. For example, while Raymond et al., (2006) only included type I eccentricity and inclination damping, Fogg \& Nelson (2007b) also included the effects of type I migration on planetary embryos in their simulations.

At the beginning of our simulations, planetary embryos are Moon- to Mars-sized (Fig. 1). How these $\sim 0.1M_{\oplus}$ bodies migrate is unclear. Real protoplanetary disks are almost certainly turbulent, and stochastic torques that result from turbulent density fluctuations can dominate the effects of type-I migration. Consequently, the orbits of low-mass planetary embryos can experience a random walk in semimajor axis (Nelson \& Papaloizou 2004; Nelson 2005) rather than a monotonic inward migration. Thus, in this work, we have performed simulations both including type I migration of embryos and neglecting this effect. For cases where type I migration of planetary embryos is taken into account simulations were performed considering the nominal type I rate  (Tanaka et al., 2002) and also a type I migration rate reduced ten times. This is achieved by multiplying ${\rm t_{m}}$, in Eq. 10, by a factor 10. Table 1 summarizes the effects that were considered for our different scenarios.

\begin{table}[h]
\scriptsize
\caption{Different models explored in our simulations concerning  the gas dissipative effects on planetesimals and protoplanetary embryos. For each model, the included forces are indicated by ($\checkmark$) and the neglected ones are shown by ($\times$)\tablenotemark{a}}
   \centering 
\begin{tabular}{@{}lcccc@{}}
  \hline
Model   &  Gas  & Radial & Ecc.  & Inc.  \\
        &  Drag (${\rm {\bold a}_{drag}}$)  & damp. (${\rm \bold{a}_m}$) & damp. (${\rm \bold{a}_e}$) &  damp. (${\rm \bold{a}_i}$)\\
  \hline\hline
Fiducial      & \checkmark    & $\times$  &  \checkmark &  \checkmark \\
\hline
I         &  \checkmark     & \checkmark  &  \checkmark &  \checkmark\\
\hline
II (Toy-Model)       &  $\times$               &  $\times$  &   $\times$ & $\times$ \\
\hline
\end{tabular}
\tablenotetext{1}{From left to right the columns represent the model, aerodynamic gas drag (${\rm {\bold a}_{drag}}$), type-I migration (${\rm {\bold a}_{m}}$), eccentricity damping (${\rm \bold{a}_e}$), and inclination damping (${\rm \bold{a}_i}$)}
\end{table}

In our fiducial model, type I migration of protoplanetary embryos is neglected. Model I includes all dissipative forces acting on the protoplanetary embryos as described before. Finally, Model II neglects all effects of the gas on planetesimals and protoplanets. Model II is clearly unrealistic but we use it to highlight the role of the gas in comparison with simulations conducted in the framework of the fiducial model.

\subsection{Migrating Super-Earths}

We consider a system of either one or multiple super-Earths originally residing beyond the outer edge of the disk of planetary embryos and planetesimals ($\geq$5AU). We considered two ``end members'' of super-Earth populations when performing our numerical simulations. First, considering a single super-Earth, we performed simulations exploring a range of masses from 5 to 15 Earth masses. However, because the evolution was quite insensitive of the super-Earth mass over the considered range, we only present results for 10 Earth-mass planets. To represent multiplanet systems of migrating super-Earths, we consider a planetary system composed by six super-Earths. The masses of these six planets were taken to be the upper limits of those of the planets in the Kepler 11 system (Lissauer et al., 2013). Table 2 shows the initial semimajor axes and masses of the migrating super-Earths in our simulations. Using  these two sets of initial conditions of super-Earths we qualitatively cover a range of possibilities within the spectrum of the planetary systems uncovered by Kepler.

\begin{table}[h]
\caption{Initial semimajor axis and mass of the migrating planets in our simulations.}
   \centering 
\begin{tabular}{@{}llcccccc@{}}
  \hline
System  & & P-I  &  P-II & P-III & P-IV & P-V & P-VI  \\
\hline 
Single  & &     &     &    &   &    &     \\
        
        & a [AU] & 5.0 &  -  &  - & - & -  & -  \\
        & M ${\rm [M_{\oplus}]}$ & 10  &  -  &  - & - & -  & -  \\
\hline
Multiplanet &  & &     &     &    &   &         \\
             & a [AU]  &  5.0           &  6.05  & 7.33  & 8.88 & 10.76  & 13.03  \\
             & M ${\rm [M_{\oplus}]}$    &  3.5            &  6  &  8 & 9.5 & 3 & 12  \\
\hline\hline
\end{tabular}
\begin{minipage}{0.45\textwidth}%
\end{minipage}%
\end{table}

Initially, the eccentricities and orbital inclinations of the super-Earths are set to zero. We apply a synthetic acceleration to force the super-Earth(s) to migrate inward on the timescale ${\rm \tau_{SE}}$\footnote{Obviously, we have always considered ${\rm \tau_{SE} \leq {\rm \tau_{diss}}}$}. This migration does not affect the evolution of the eccentricity and orbital inclination of the migrating planets.  To save computational time when running our simulations, we removed the migrating super-Earths from the system when they reached a heliocentric distance equal to 0.1 AU. This assumption does not bias our results, since we are interested in the accretion of terrestrial planets from the objects left beyond 0.5 AU after super-Earth(s) migration and gas disk dissipation.

For migrating multi-super-Earth systems, we tested two different migration scenarios. The first scenario -- called ``locked migration'' -- mimics inward migration in a stable multiresonant configuration as observed in hydrodynamical simulations (Terquem \& Papaloizou, 2007; Cresswell \& Nelson 2008; Morbidelli et al., 2008; Horn et al., 2012; Pierens et al., 2013). The period ratios of the super-Earths are kept fixed during the migration and, to ensure the stability of the system, we neglect the mutual interactions among the super-Earths.

The second scenario of migration for the multiplanet system of super-Earths is called  ``migrating in sequence''. In this scenario, we consider that the super-Earths migrate one by one. Each one of them migrates on a timescale ${\rm \simeq\tau_{SE}/n}$, where n  is the number of migrating super-Earths in the system.  Numbering the super-Earths according to their initial heliocentric distance, where the closest one is labeled by ``1'' and the farthest by ``n'' (in our case n=6), the ``${\rm ith}$'' super-Earth is only allowed to migrate when the ``${\rm (i-1)th}$''  super-Earth has reached a  heliocentric distance of 0.1 AU, and, consequently, has been removed from the system. Table 3 shows the simulations that were performed considering different scenarios of migrating super-Earths and the models for the gas effects  defined in Table 1.

\begin{table}[!h]
\caption{Summary of simulations performed considering different configurations of migrating super-Earths combined with gas effect models of Table 1.}
   \centering 
\begin{tabular}{@{}lccc@{}}
  \hline
Config.  &  Fiducial   & Model I & Model II  \\
Single              &  \checkmark$^\star$  & $\times$ & \checkmark\\
Multiplanet/locked        &  \checkmark & \checkmark & \checkmark\\
Multiplanet/sequence  &  \checkmark  & $\times$ &  \checkmark \\
 \hline
\end{tabular}
\begin{minipage}{0.45\textwidth}%
\tiny Note: The ``$\star$'' shows the case for which we have also carried simulations considering different size for the planetesimals, i.e. 10 and 1000km, the fiducial size being equal to 100 km.
\end{minipage}%
\end{table}

\subsubsection{Super Earths Migration Timescale}

A key issue is the effect of the super-Earths' migration timescale ${\rm \tau_{SE}}$. As discussed above, the rate of type I migration is an open issue. Super-Earths may  migrate either fast or slowly depending on the gas-disk structure and thermodynamics (see discussion in Section 1.1). For this reason, we have adopted two migration modes for the super-Earths in our simulations (Table 2), which we called of ``fast'' and ``slow'' migrating scenarios.  Our fast inward migration is characterized by the timescale estimated in isothermal disk models for a 10 Earth-mass planet to migrate from 5 AU down to the star: 0.01-0.1 Myr (Ward, 1997). On the other hand, our slow-migrating planet(s) have  migration timescales comparable to the gas-disk lifetime (Lyra et al., 2010; Horn et al., 2012), which we considered to range between 1 and 10 Myr.

\subsubsection{Prescription for Super-Earth's migration} 

In our simulations, the super-Earths' migration is implemented by applying an external synthetic torque to each planet's orbit such that its semimajor axis  varies as
 
\begin{equation}
{\rm a_p(t) \simeq a_{ini} + \Delta a\frac{t-t_{ini}}{\tau_{SE}}}
\end{equation} 
where  ${\rm a_{ini}}$ is the initial semimajor axis of the planet, ${\rm \Delta a}$ is the radial displacement, t is the time, ${\rm t_{ini}}$ is the time that each planet starts to migrate and  ${\rm \tau_{SE}}$ is the timescale for planet migration as mentioned before. The ${\rm \Delta a}$ value of each migrating super-Earth is purposely chosen  such that the inward-migrating planet comes sufficiently close to the star to be artificially removed from the simulation (according to our criterion of removal) on the timescale ${\rm \simeq\tau_{SE}}$.

\section{Results}

As a super-Earth migrates through a sea of planetesimals and protoplanetary embryos, any given small body will follow one of three possible evolutionary pathways. Planetary bodies are either shepherded, scattered into external orbits, or accreted by super-Earths during their passage (Ward \& Hahn, 1995; Tanaka \& Ida, 1996; 1997). No planetary embryos or planetesimals were ejected in any of our simulations, simply because the relative velocities between the super-Earths and these objects are too small to scatter objects strongly enough (e.g., Ford \& Rasio 2008).

When a super-Earth interacts with the disk of planetesimals and planetary embryos it stir ups these objects and increases their velocity dispersion by gravitational scattering (Lissauer, 1987). The protoplanetary objects recoil in semimajor axis away from the super-Earth while they are scattered into more eccentric and inclined orbits. If an efficient dissipative force acts at the same time on the protoplanetary objects, such as aerodynamic gas drag, dynamical friction or tidal damping from the disk of gas, their orbits tend to be circularized. As a result, these objects are driven inward, ahead of the migrating super-Earth, which results in an effective shepherding process. If, instead, there are no efficient dissipative forces acting on the protoplanetary bodies, the latter have their orbits perturbed until they eventually cross the orbit of the super-Earth. This favors collisions with the super-Earth  or scattering into external orbits (Tanaka \& Ida, 1999; Mandell et al 2007). 

Below we detail the evolution of the system in these various scenarios (as summarized in Tables 1, 2 and 3).

\subsection{A Single Migrating Super-Earth}

We first present the results of simulations with a single migrating super-Earth. The super-Earth is initially placed at 5 AU (see Table 2) and migrates inward at the start of the simulations. The semimajor axis of the super-Earth varies as in Eq (17) owing to the synthetic forces applied to its orbit. 

\subsubsection{Dependence of the Results on the Migration Speed}

Figures 2 and 3 show the dynamical evolution of two systems containing a single 10 Earth-mass planet migrating on different timescales. Both simulations were performed in the framework of our fiducial model (Table 1).

\begin{figure*}
\begin{minipage}{\textwidth}
\centering
\includegraphics[scale=.63]{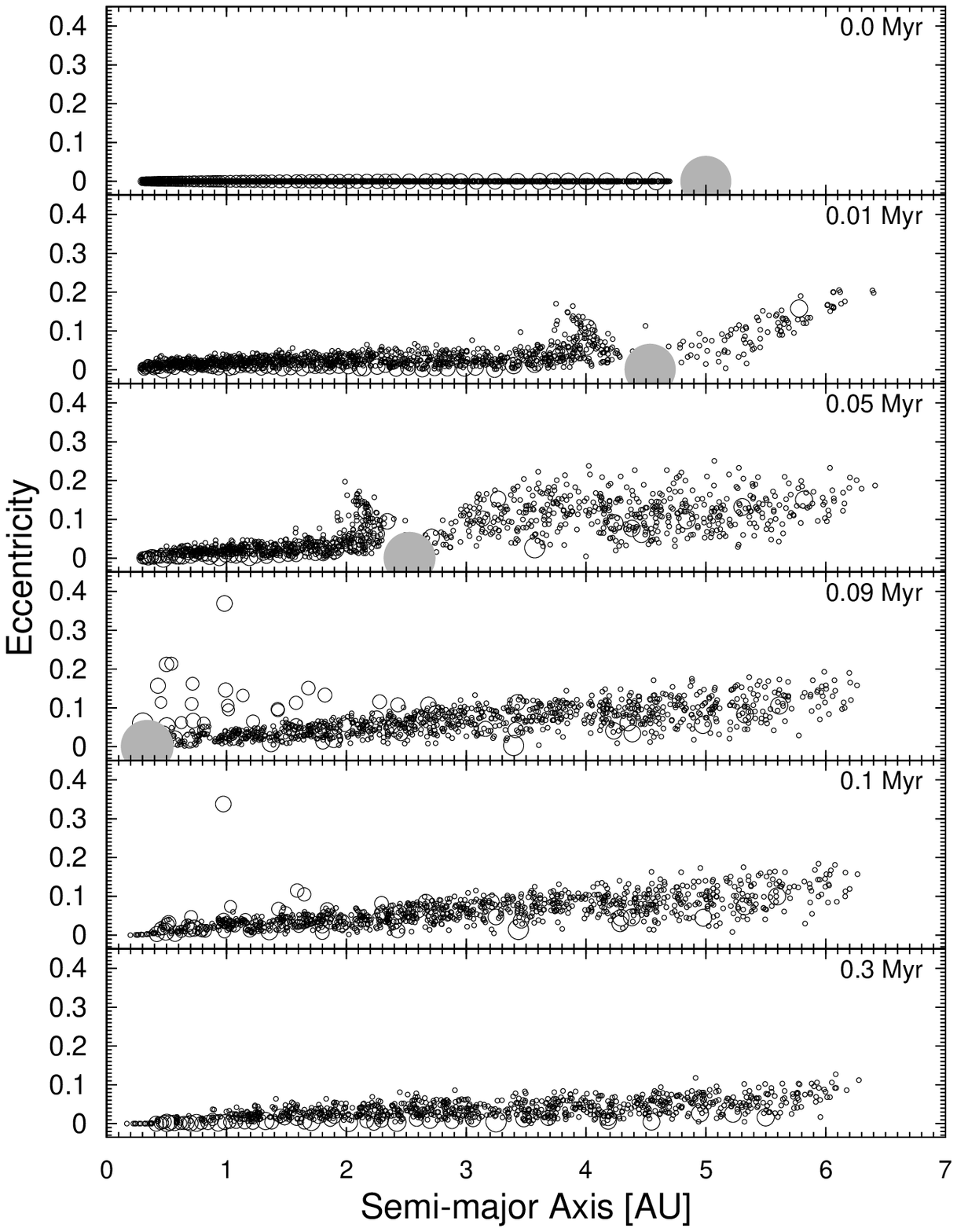}
\includegraphics[scale=.63]{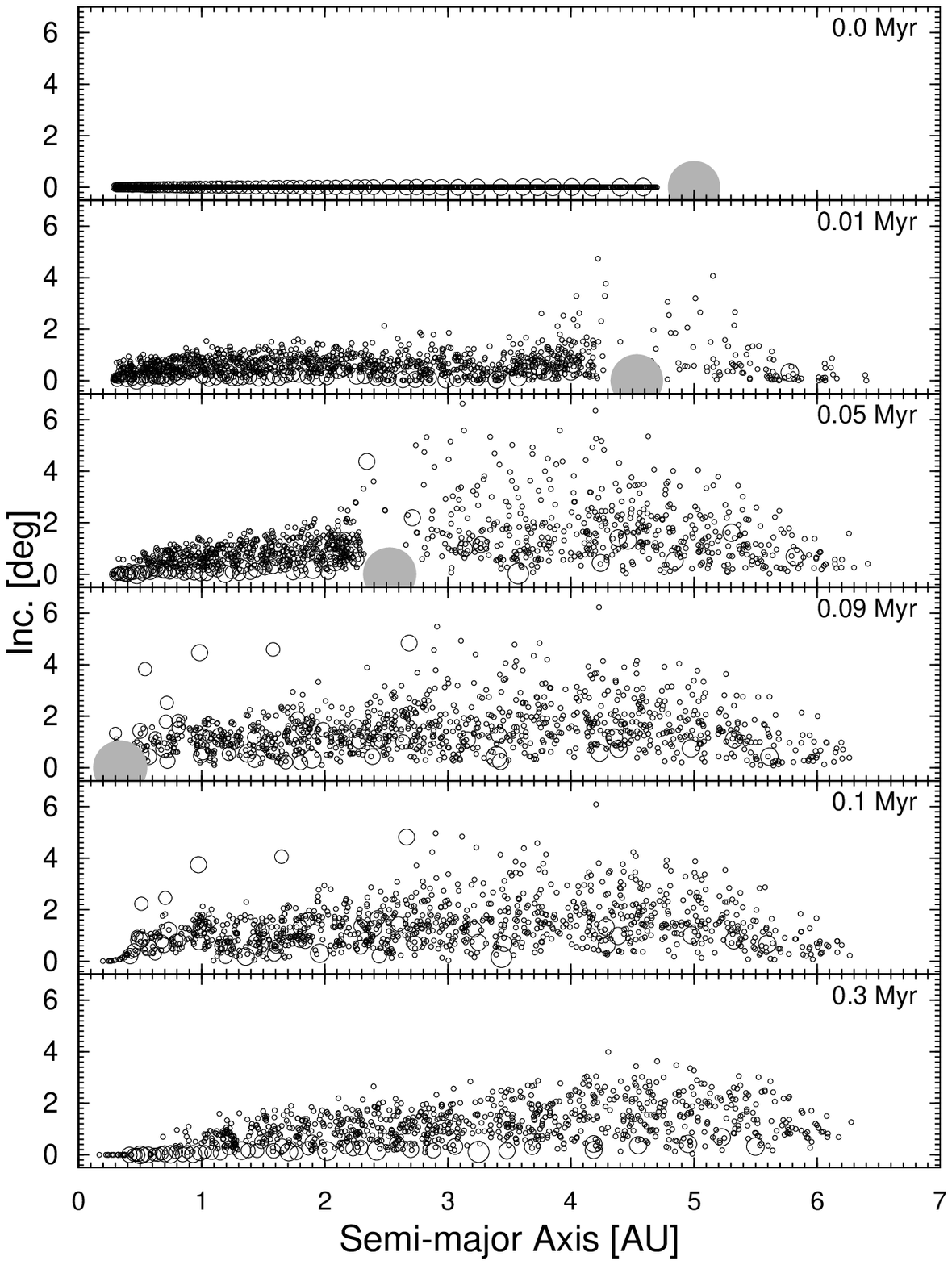}
\caption{Snapshots of the dynamical evolution of a disk of planetesimals and protoplanetary embryos in presence of a single migrating Super-Earth with mass equal to ${\rm 10M_{\oplus}}$. The gas disk lifetime is  ${\rm \tau_{gas}=10Myr}$, and the super-Earth migrates on a timescale ${\rm \tau_{SE}= 0.1 Myr}$. The left-hand  panel shows the dynamical evolution in a diagram semimajor axis versus eccentricity while the right-hand panel shows the dynamical evolution in a diagram semimajor axis versus inclination.  The size of each dot is proportional to ${\rm M^{1/3}}$, where ${\rm M}$ is the mass of the corresponding body.} 
\end{minipage}
\end{figure*}

Figure 2 shows the evolution of a simulation in which a super-Earth migrated inward in 0.1 Myr. The super-Earth rapidly opens a gap around its orbit by scattering (or accreting) nearby planetesimals and protoplanetary embryos (Tanaka \& Ida, 1997).  Within 0.01 Myr, the super-Earth has moved from 5 to 4.5 AU and the signature of scattering of nearby bodies is apparent. As the super-Earth migrates through the sea of planetesimals and embryos, it scatters most of the bodies it encounters onto external orbits rather than shepherding them along its own radial excursion. Consequently, at the end of the migration, the protoplanetary disk is only weakly perturbed by the passage of the super-Earth and a significant portion of the disk's initial mass survives the super-Earth's migration.  Only about 5-10\% of the protoplanetary bodies collide with the super-Earth instead of being scattered. An additional $\sim$ 10-15\% of planetary embryos and planetesimals fall into the star. 

There are two explanations for the survival of most of the disk. First, the super-Earth is not very massive.  A Jupiter mass planet migrating at this same migration rate would produce a very different outcome (Raymond et al., 2006). Second, the migration is fast.  The gas drag and tidal forces acting on planetesimals and protoplanetary embryos do not have enough time to damp the eccentricities of the excited bodies, so shepherding is inefficient. 

Although most of bodies were scattered during the super-Earth's passage, the majority of the disk's solid mass remains confined within the original region. In fact, ${\rm \sim 75\%}$ of the initial mass of embryos and planetesimals survives inside 5 AU at the end of the simulation (Figures 2 and 4). This is also a consequence of the fast migration speed of the super-Earth, for two reasons. First, a  fast migration speed implies that, for each object, the scattering phase is short; thus no object can be scattered onto a distant orbit. Second, a fast migration inhibits the shepherding process, as stated above.

Although Figure 2 shows the dynamical evolution of the system only until 0.3 Myr, we numerically integrated the evolution of the system for another 10 Myr. This second part of the evolution follows a pattern that is well documented in the literature. While the gaseous disk is still present, planetesimals and embryos stay in a dynamically cold orbital configuration.  This promotes the rapid growth of the embryos (Kokubo \& Ida, 2000). After gas dispersal, planetary embryos stir up planetesimals and increase their velocity dispersion, which drastically slows their own growth rate (Ida \& Makino, 1993). The system of embryos eventually becomes unstable and giant collisions start, leading to the formation of a terrestrial planets in stable orbits (Chambers \& Wetherill, 1998). In figure 2, it is evident that the emergence  of terrestrial-like planets is likely to happen given that, after the super-Earth migration and gas dissipation, the mass between 0.7 and 1.5 AU is about ${\rm \sim 1 M_{\oplus}}$. 

\begin{figure*}
\begin{minipage}{\textwidth}
\includegraphics[scale=.63]{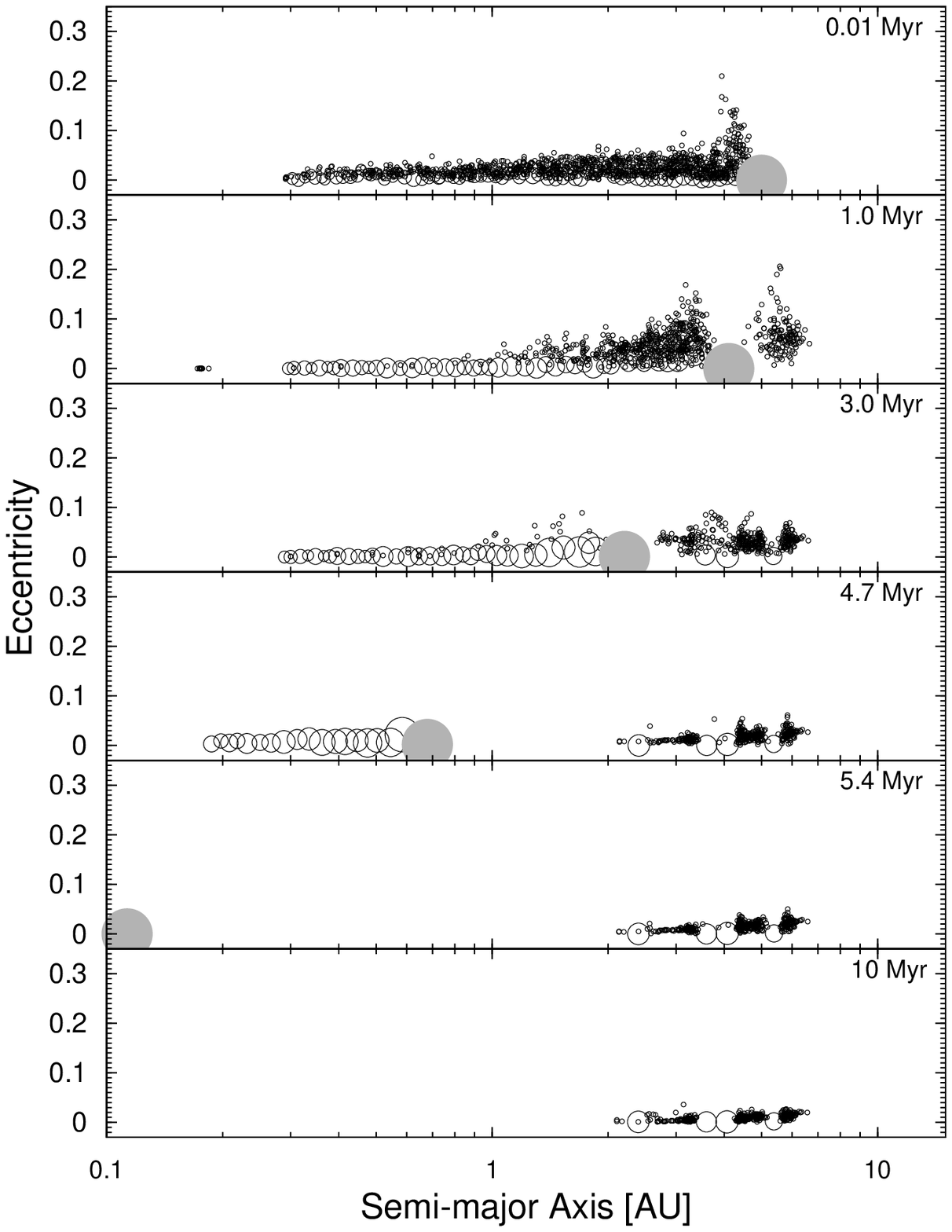}
\includegraphics[scale=.63]{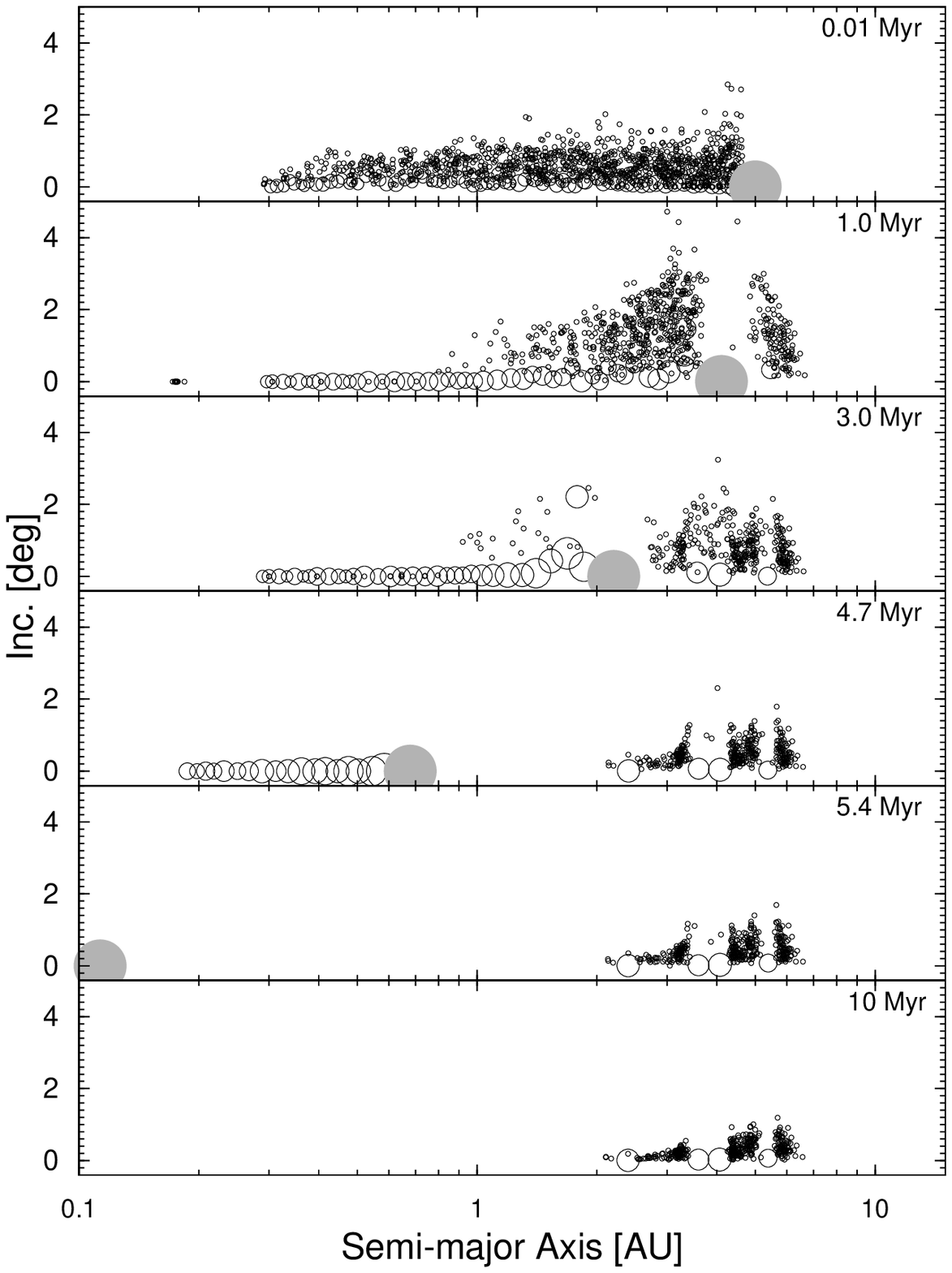}
\caption{Same as Fig. 2, but for a case where ${\rm \tau_{gas}=10Myr}$, and ${\rm \tau_{SE}= 5 Myr}$. }
\end{minipage}
\end{figure*}
 
Figure 3 shows a contrasting scenario with a slowly migrating super-Earth.  The migration timescale ${\rm \tau_{SE}}$ is 5 Myr. This slow migrating mode results in trends significantly different from those in Figure 2. First, before the super-Earth reaches the terrestrial planet region the number of planetesimals has decreased drastically owing to accretion by protoplanetary embryos.  Second, most embryos are shepherded inward by the super-Earth instead of being scattered onto external orbits. This is because the slow migration rate of the super-Earth, combined with the strong gas dissipative effects on the embryos, favors capture into mean motion resonances between the super-Earth and the embryos.  Long chains of embryos in mean motion resonance are maintained until the bodies reach a heliocentric distance of 0.1 AU and are removed from the simulation. In this case the amount of leftover material in the terrestrial zone after the passage of the super-Earths is likely insufficient to build a 1 Earth-mass planet around 1 AU.

Figure 4 shows the final mass distribution of protoplanetary embryos and planetesimals in the disk after the super-Earth passage for simulations considering different timescales for super-Earth's migration. For all cases, the mass distribution is computed at 10 Myr. For each scenario, the results of 3 or more different simulations with slightly different initial conditions for planetary embryos and planetesimals are shown (see Section 2.1). When considering the fiducial model, a migration timescale longer than 1 Myr results in an effective shepherding process. For these cases, the final amount of mass between 0.7 and 1.5 AU is, in general, much smaller than 1 Earth mass, if not equal to zero.

\begin{figure*}
\begin{minipage}{\textwidth}
\includegraphics[scale=.8]{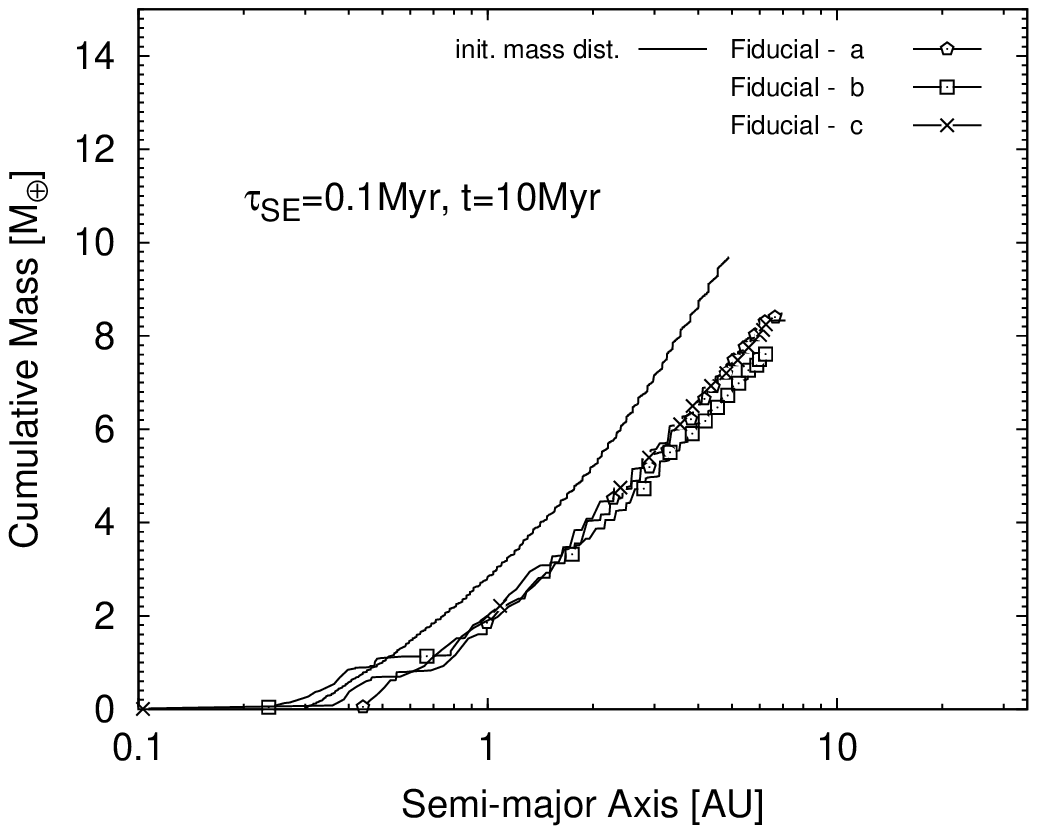}
\includegraphics[scale=.8]{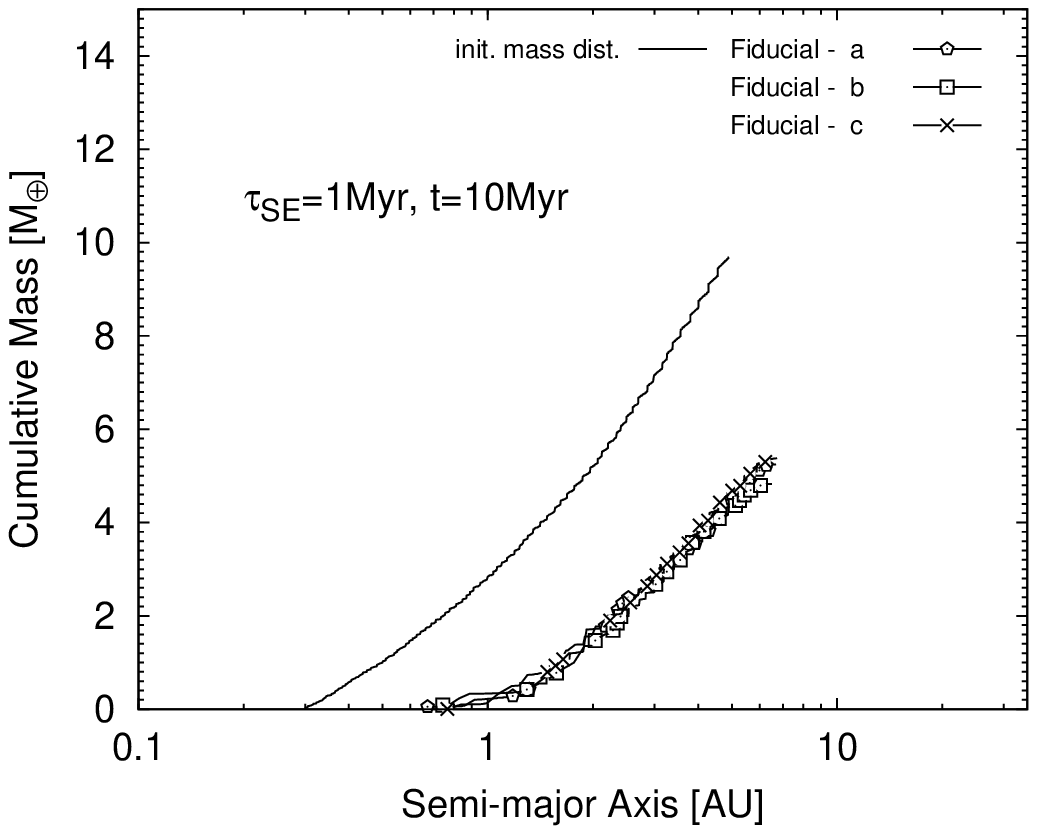}

\includegraphics[scale=.8]{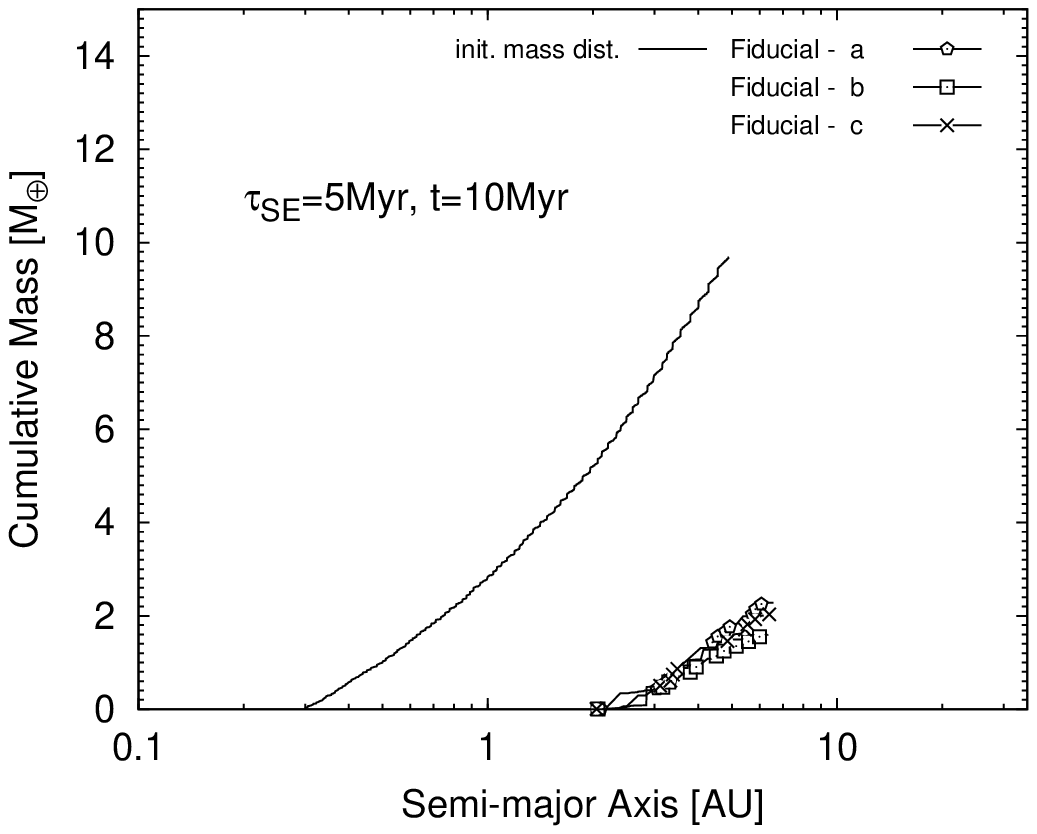}
\includegraphics[scale=.8]{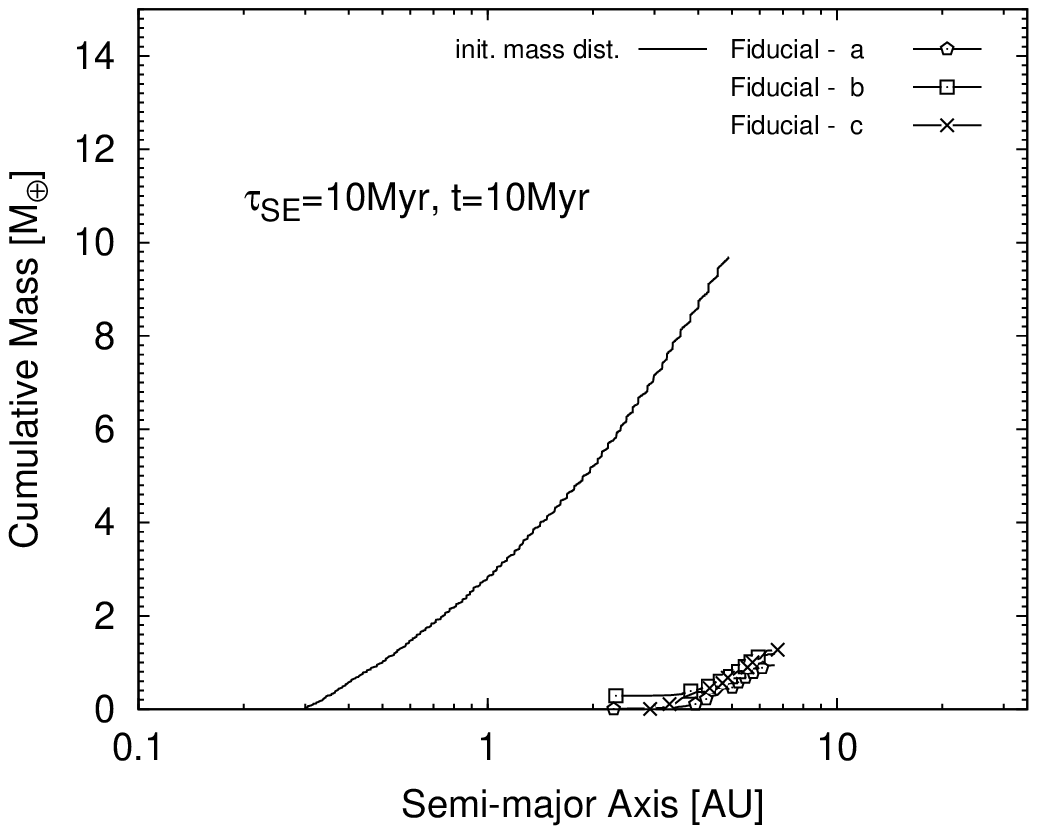}
\caption{Cumulative mass distribution vs semimajor axis in the protoplanetary disk, after the super-Earth's passage, for different migration speeds. All these results corresponds to simulations considering a single migrating super-Earth. The solid line represents the original mass distribution in protoplanetary embryos and planetesimals. Solid lines overlapped by points represent simulations
conducted in the framework  of our fiducial model. The point styles on the lines represent three different simulations (a, b and c) with only slightly different initial condition for the planetary embryos and planetesimals. The time at which the cumulative mass distribution is computed is 10 Myr.}
\end{minipage}
\end{figure*}

\subsubsection{Gas Effects}

We have also analyzed how the dissipation timescale of the gaseous disk affects the final results. Our analysis is made by comparing the results of the simulations of Figure 4, where $\tau_{gas}=10$~Myr, with simulations enacting the same super-Earth migration rates but using smaller values for $\tau_{gas}$.  These simulations were in the framework of our fiducial model.

\begin{figure*}
\begin{minipage}{\textwidth}
\includegraphics[scale=.8]{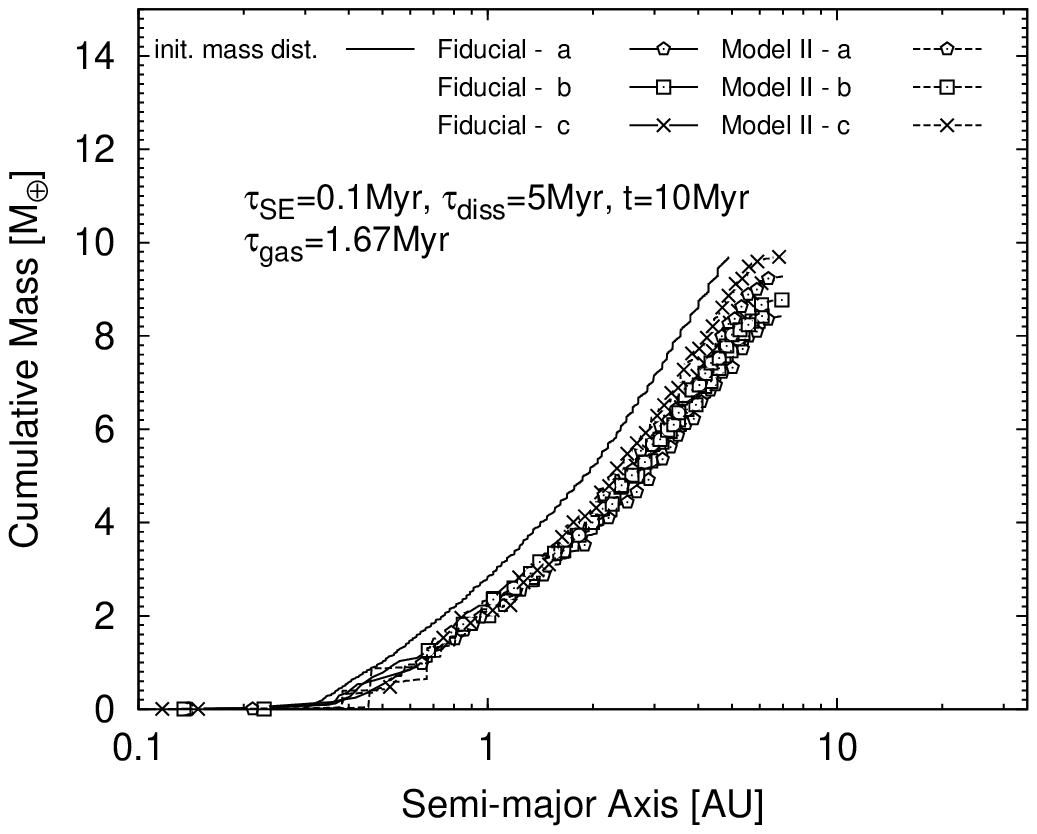}
\includegraphics[scale=.8]{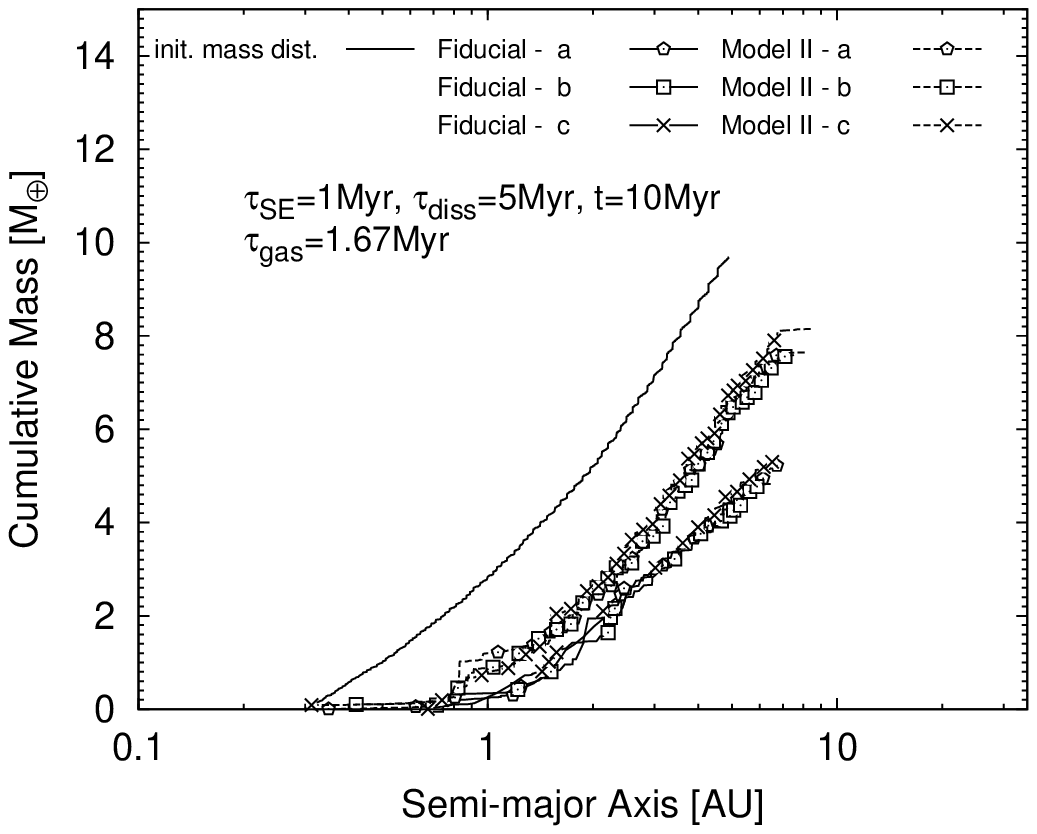}

\includegraphics[scale=.8]{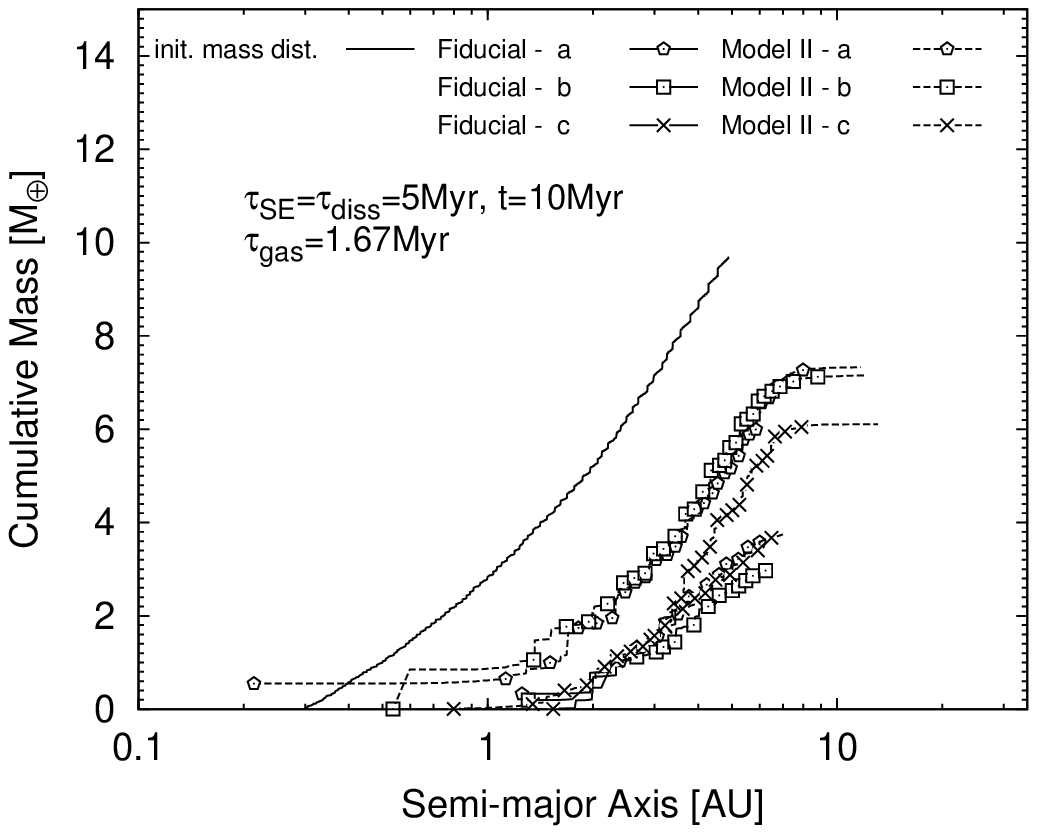}
\includegraphics[scale=.8]{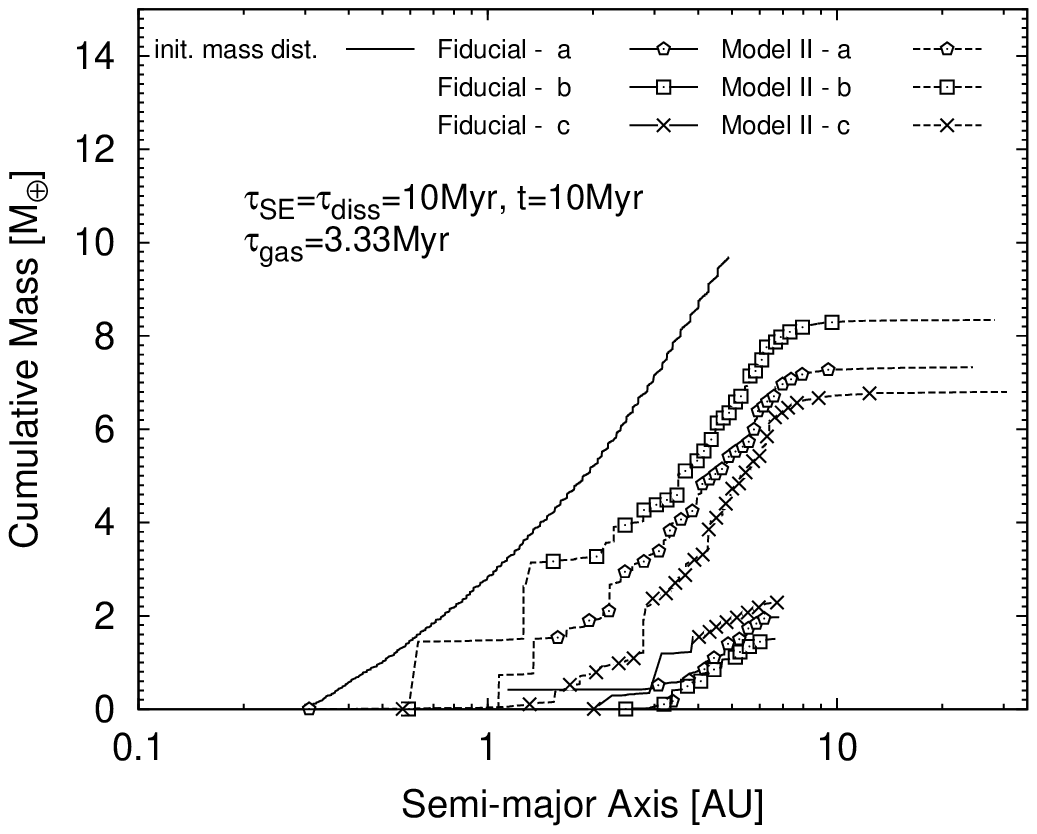}
\caption{Same as Fig. 4 but for simulations with different gas dissipations timescales. Solid lines overlapped by points represent simulations with the value of $\tau_{gas}$ reported in the panel, while dashed lines overlapped by points correspond to the results obtained within Model II, i.e. neglecting all damping effects of the gas. The values of  $\tau_{diss}$ are also shown.}
\end{minipage}
\end{figure*}

Figure 5 shows the results for the faster dissipation timescales. Again, the cumulative mass is shown for all cases at 10 Myr. Comparing Figure 4 and Figure 5 we note that different dissipation timescales do not change the results dramatically. However, looking in detail we see that in those simulations where the gas disk dissipates faster, the final disk of planetesimals and embryos tends to be slightly less depleted.

A more dramatic way to highlight the effects of gas is to compare with simulations that neglect the gas-damping effects on planetesimals and protoplanetary embryos (Model II). The resulting mass distributions in the disk are also shown in Figure 5.  For fast-migrating super-Earths ($\tau_{SE}\le 1$~Myr) there is little difference between the simulations with and without gas drag.  Gravitational scattering is the dominant process in these systems.  But for a slowly migrating super-Earth gas drag plays a very important role.  If gas drag is neglected a much larger fraction of terrestrial material survives at 1 AU than when gas drag is included.  Accounting for the gas dissipative effects on the protoplanetary embryos (embryos and planetesimals) is crucial for the realism of the simulations.

\subsection{Multiplanet Configuration of Migrating Super-Earths}

In this section we present the results of simulations considering a multiplanet configuration of migrating super-Earths.  We include six super-Earths with masses similar to those of the planets in the Kepler 11 system (Lissauer et al., 2011, 2013). The exact locations of these planets at the beginning of our simulations are shown in Table 2.  We explored many different scenarios, considering different migration timescales for these bodies, gaseous disk lifetimes, and two different arrangements for the migrating bodies, as defined in section 3.2. The results of simulations considering the scenario ``migration in sequence''  will be presented in Section 4.2.3.

\subsubsection{Migration Speed}

When considering a convoy of migrating super-Earths, the migration speed has the same effect as for a single super-Earth.  That is, when the super-Earths migrate fast, on a timescale $\sim$0.1~Myr or shorter, the protoplanetary disk is only weakly disturbed and a significant fraction of the initial mass survives after the migration. Figure 6 shows the dynamical evolution of 6 super-Earths migrating together (in a locked configuration) on a timescale of 0.1~Myr.  Of course, the protoplanetary disk is more disturbed than when a single super-Earth is considered. However, the system is quickly cooled down owing to the dissipative effects of the gas exerted on planetary embryos and planetesimals after the passage of the super-Earths.

Figure 7 shows  a simulation of a system of slowly migrating super-Earths ($\tau_{SE} = 5$~Myr).  As expected, shepherding dominates over scattering, as was the case for a single super-Earth.  After the super-Earths migration, the region between 0.5 and 3 AU is virtually devoid of solid material.

Figure 8 shows the final cumulative mass distributions in the disk of planetesimals and planetary embryos for simulations for different super-Earth migration speeds. These simulations are similar to those in Figure 4 for the fiducial model with a single super-Earth. However, as expected, less material survives around 1 AU for the migrating multiple-super-Earth system, reducing the probability of forming terrestrial planets (or perhaps just producing lower-mass terrestrial planets; Raymond et al 2007b).  Given the stronger scattering in the multiplanet scenario, some planetesimals and embryos survive on distant orbits, with semimajor axes up to 14 AU.  In contrast, nothing was scattered past 7-8 AU in the simulations with a single super-Earth migrating at the same speed (compare Figure 8 with Figure 4).

\begin{figure*}
\begin{minipage}{\textwidth}
\includegraphics[scale=.63]{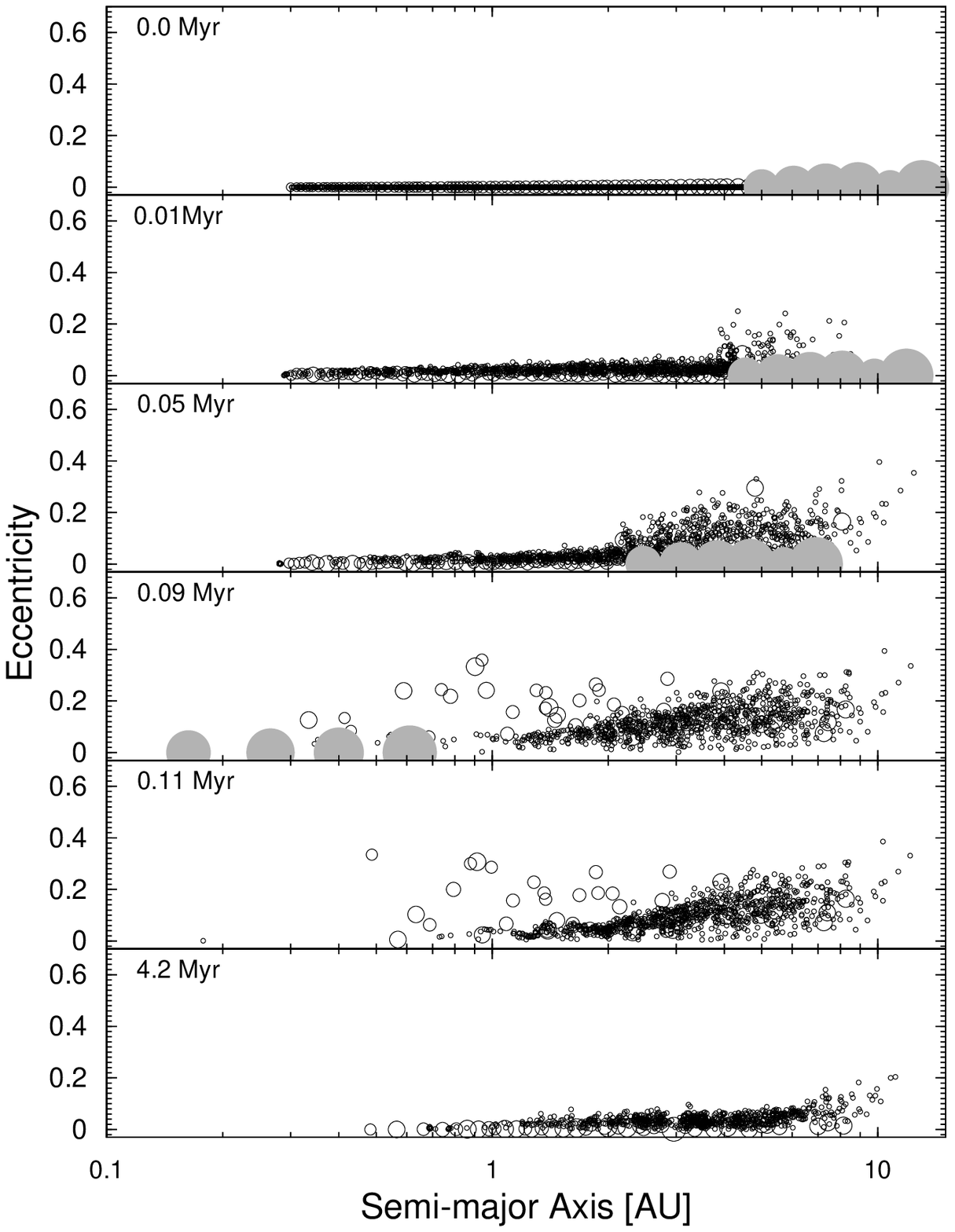}
\includegraphics[scale=.63]{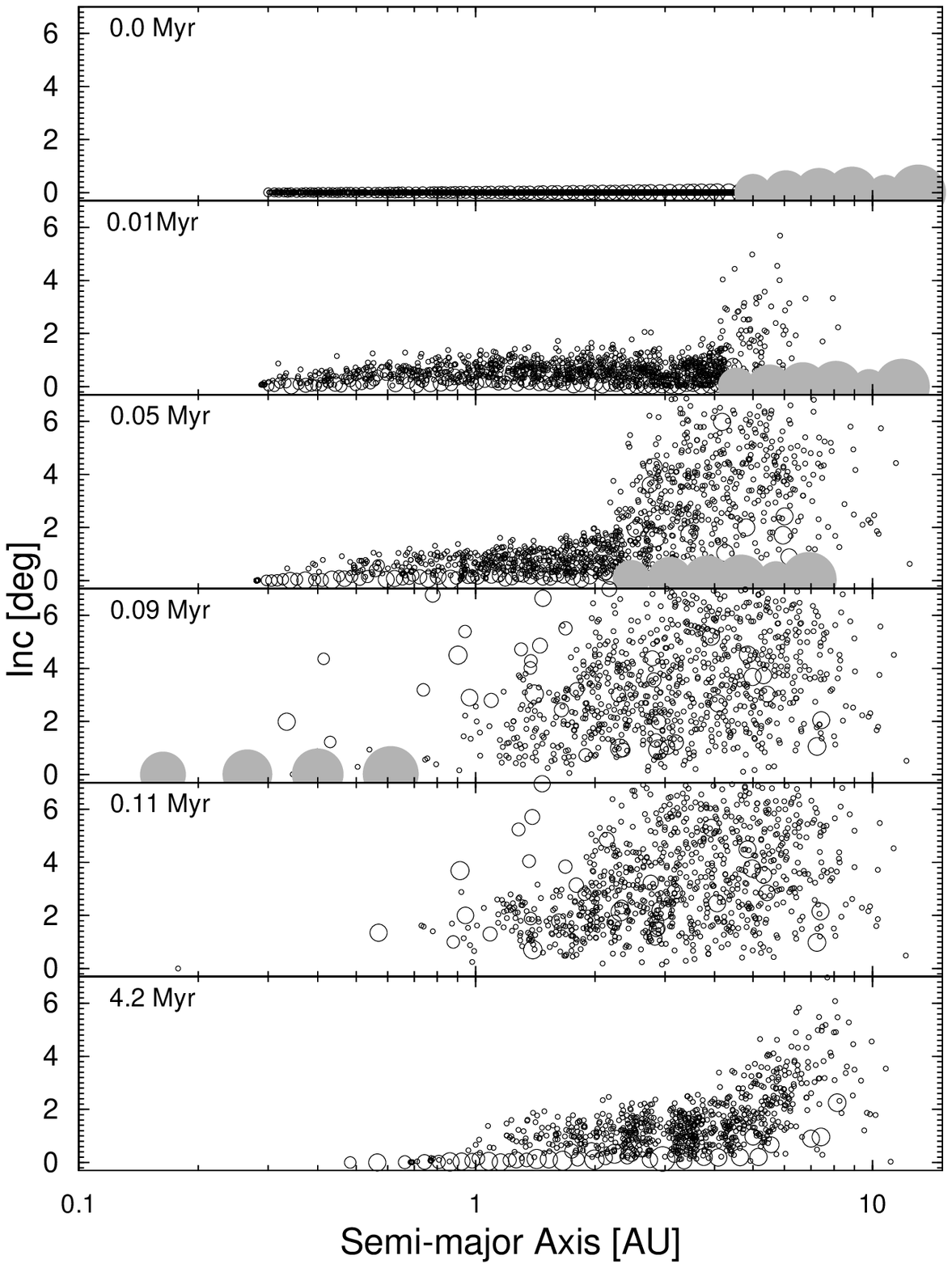}
\caption{Same as Figure 3, but for a system of 6 super-Earths.  The gas disk lifetime is ${\rm \tau_{gas}=10Myr}$, and the super-Earths migrate on a timescale ${\rm \tau_{SE}= 0.1 Myr}$ in a locked-configuration.}
\end{minipage}
\end{figure*}

\begin{figure*}
\begin{minipage}{\textwidth}
\includegraphics[scale=.63]{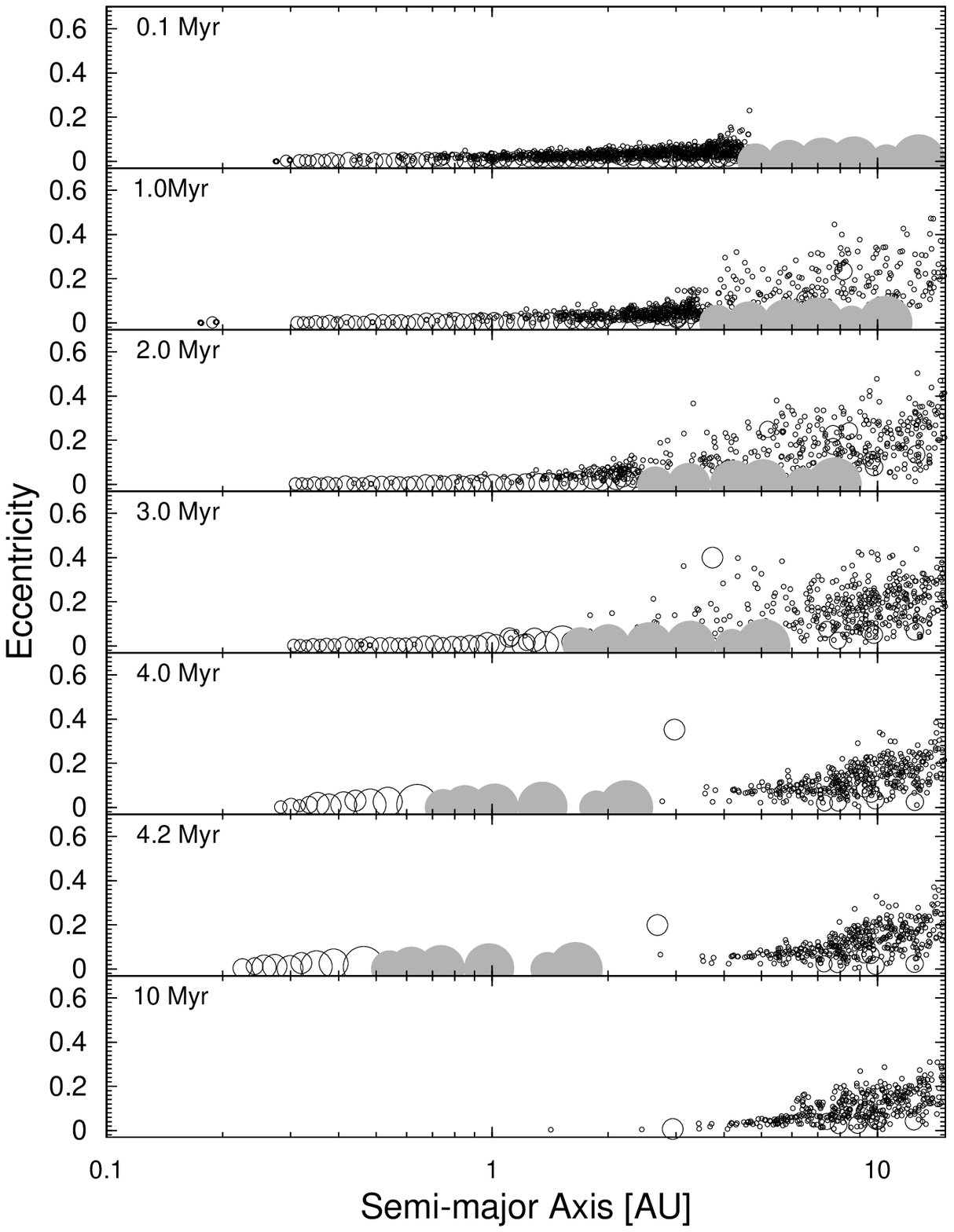}
\includegraphics[scale=.63]{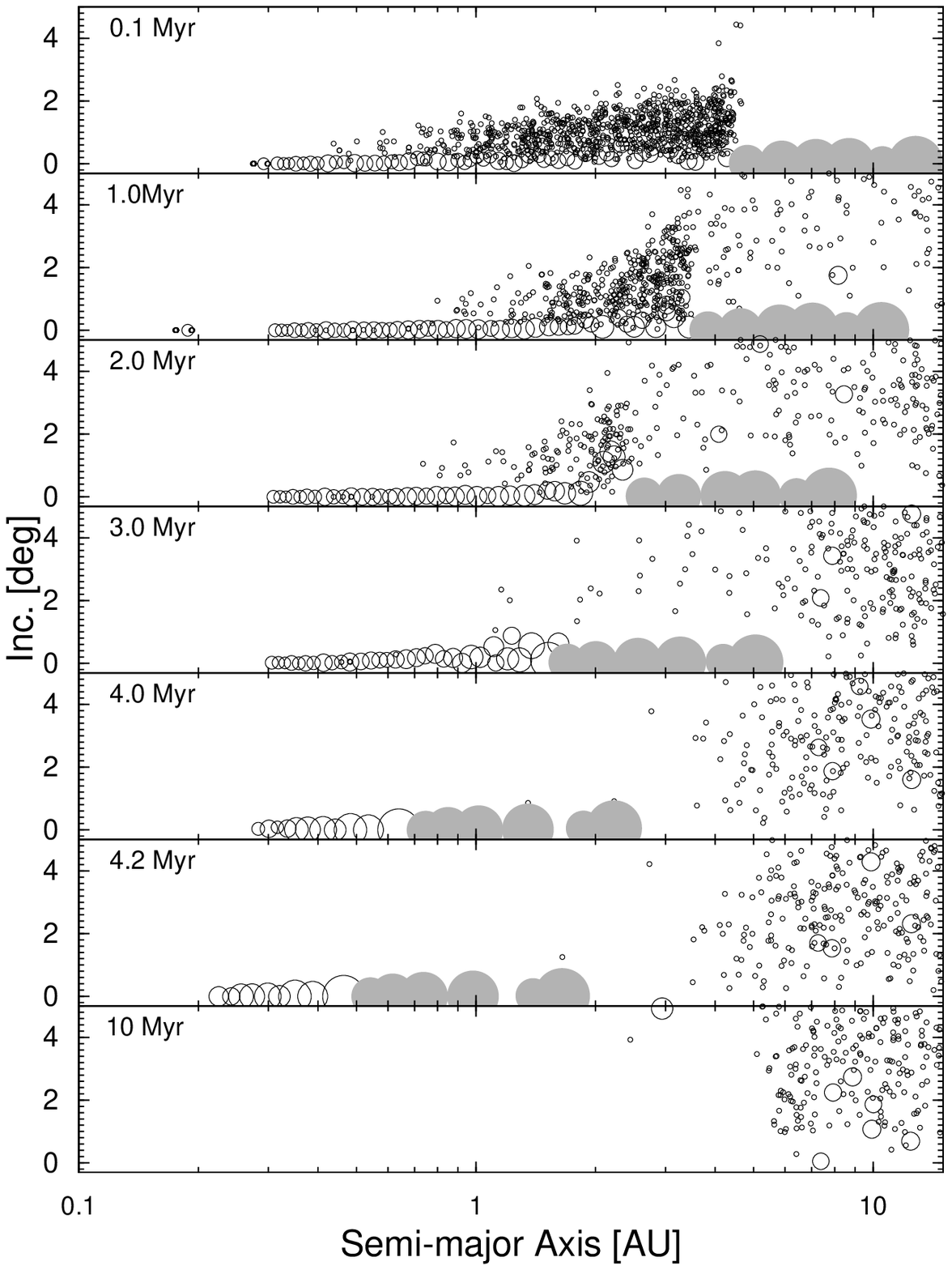}
\caption{Same as figure 6, but for ${\rm \tau_{SE}= 5 Myr}$} 
\end{minipage}
\end{figure*}

\begin{figure*}
\begin{minipage}{\textwidth}
\includegraphics[scale=.8]{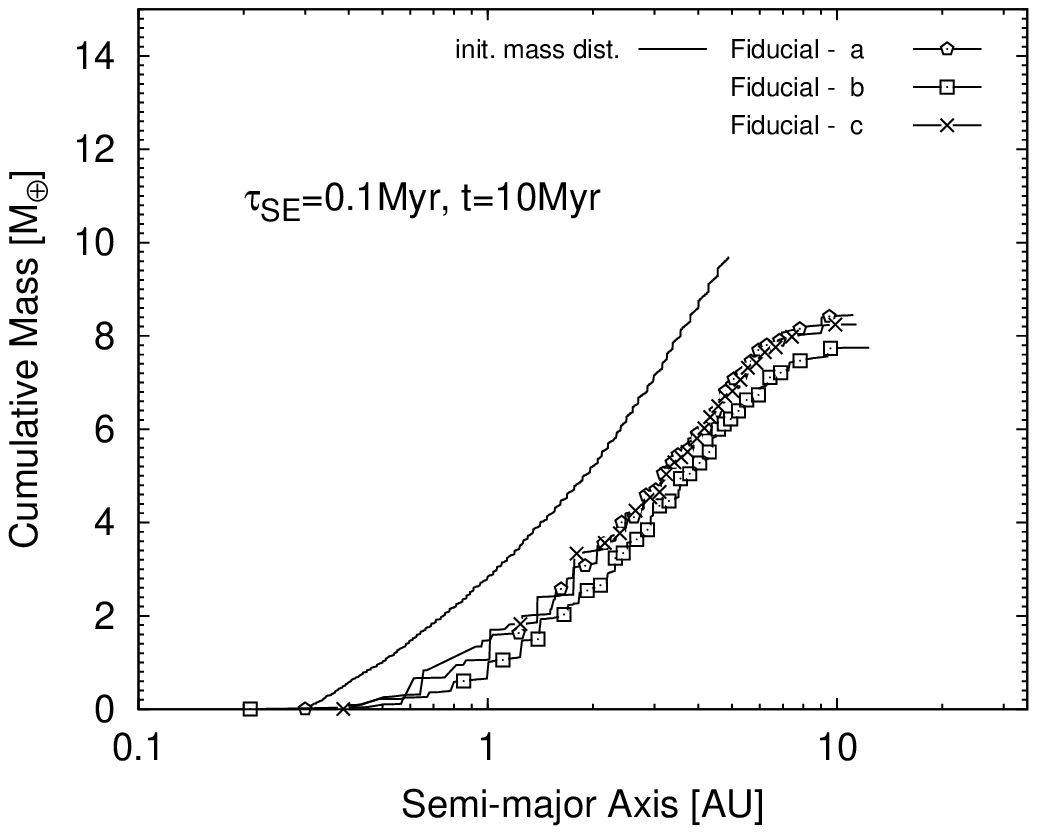}
\includegraphics[scale=.8]{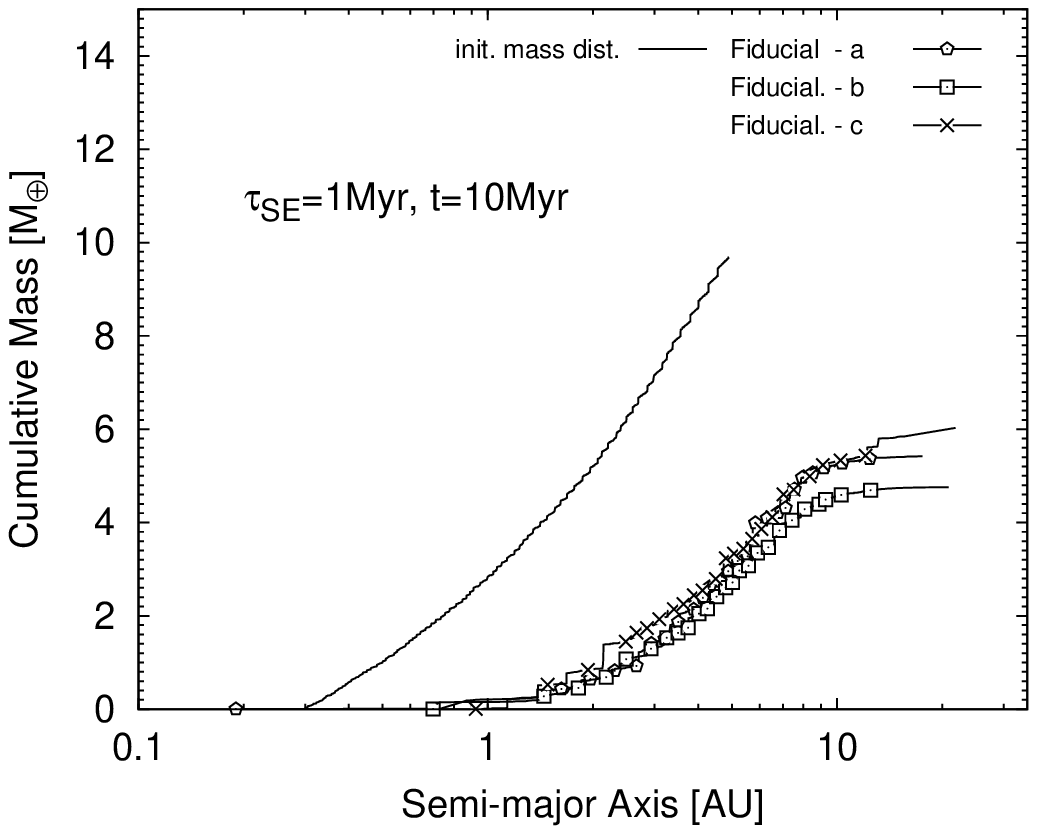}

\includegraphics[scale=.8]{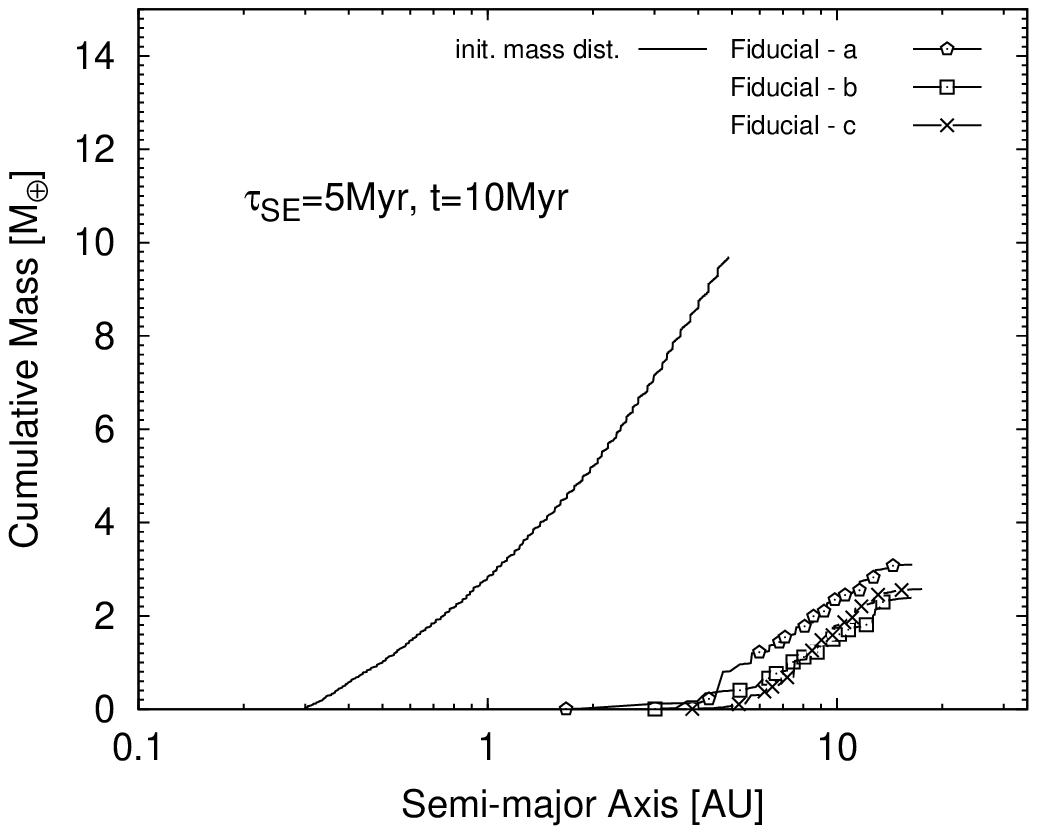}
\includegraphics[scale=.8]{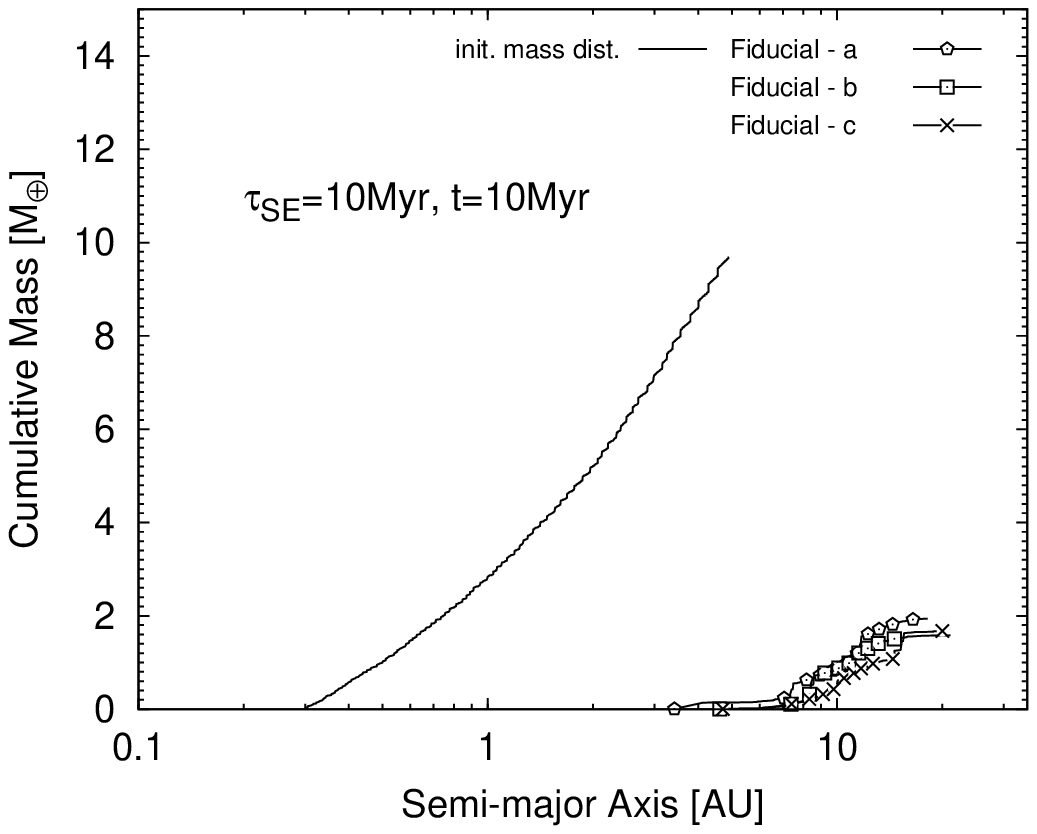}
\caption{Same as Figure 4 but for a system of six super-Earths with different migration timescales, reported on each panel. The super-Earths migrate in a locked configuration.}
\end{minipage}
\end{figure*}

\subsubsection{Dependence on planetesimal size and type-I migration of planetary embryos}

\paragraph{Planetesimal Size} We ran a suite of simulations all adopting the same timescales for the super-Earth's migration and for the gas disk dissipation.  We find that that the planetesimal size does not significantly change the trends observed before for planetesimals sizes of 100~km. 

Figure 9 shows a comparison between the final cumulative mass distributions in simulations with planetesimal sizes of  10 and 1000~Km. This comparison is done considering the multiple-super-Earth system, for two different migration timescales. Simulations with smaller planetesimals tend to produce final protoplanetary disks with a smaller amount of surviving mass. This happens for two reasons. Smaller planetesimals undergo faster radial drift than bigger ones, so many spiral onto the star before the super-Earth's passage. Second, when planetesimals are smaller their orbits are more effectively damped by gas drag and therefore they are more efficiently accreted by the protoplanetary embryos, which are then pushed toward the star by the migrating super-Earths.

\begin{figure}
\includegraphics[scale=.8]{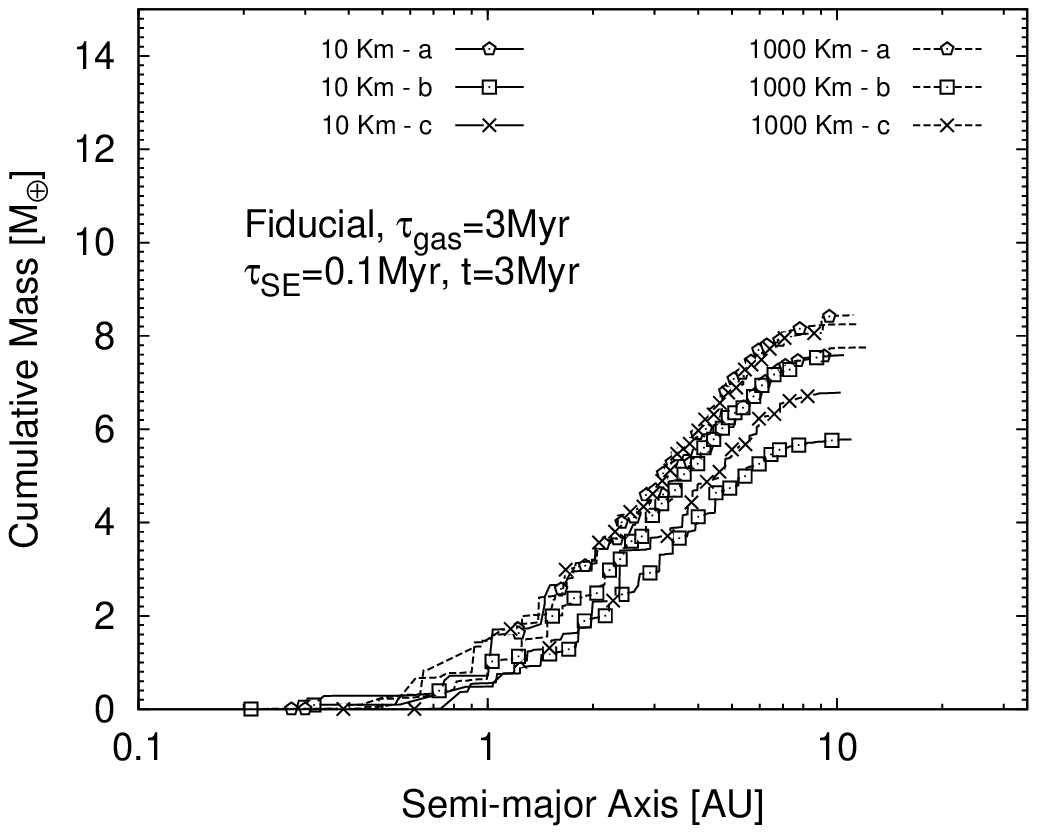}
\includegraphics[scale=.8]{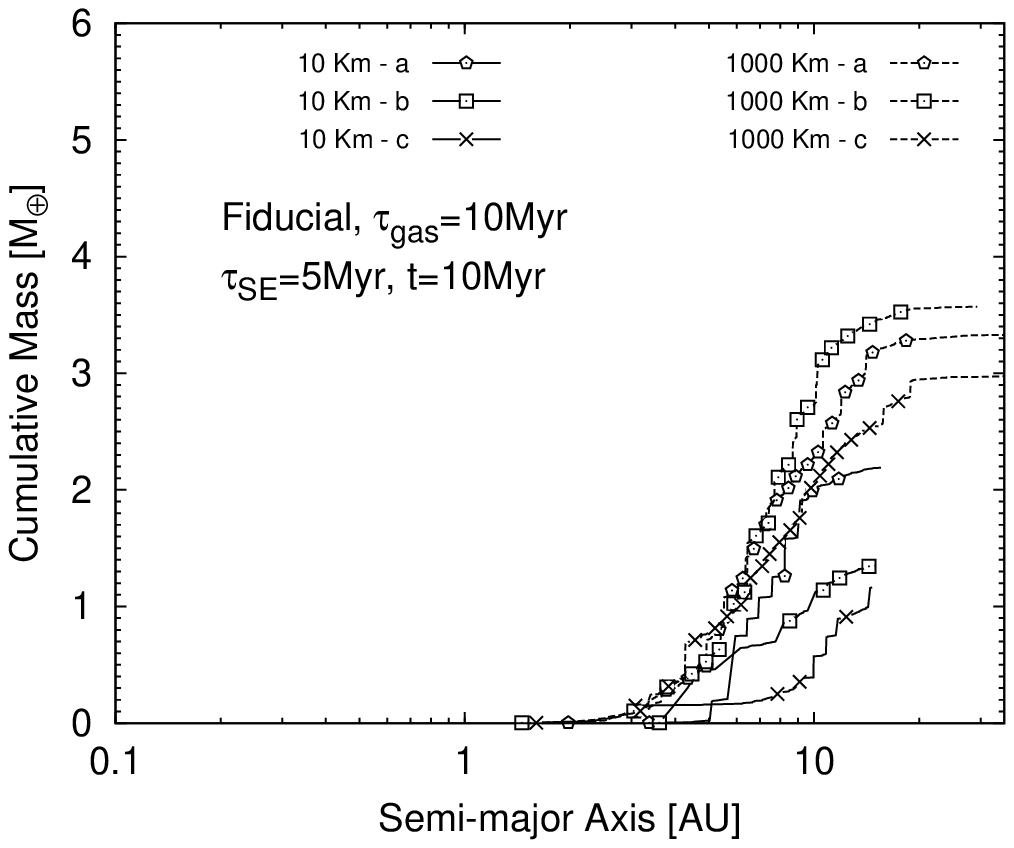}
\caption{Same as Figure 4, for different planetesimal sizes and two migration timescales for super-Earths.}
\end{figure}

\paragraph{Effects of type-I migration of planetary embryos}

We also performed simulations including type I migration of planetary embryos. These experiments correspond to Model I (Table 1). Type I migration of planetary embryos was taken into account in two ways. In the first series of simulations we imposed the nominal type I migration rate, given by Eq. 10. In the second series, we considered a reduced type-I migration rate, 10 times slower than the nominal one. Recall that our fiducial model neglects type-I migration of the planetary embryos (i.e. ${\rm t_m=\infty}$ in Eq.~10). 

Representative results of these simulations are shown in Figure 10, again in the form of the final cumulative mass distributions in the disk of embryos and planetesimals. The two panels correspond to different migration rates for the super-Earths and dissipation timescales for the gas.  In general, the simulations implementing type I migration of the embryos present trends similar to those of our fiducial model. However, qualitatively, the faster the type I migration of the embryos the smaller the final mass in the protoplanetary disk. In order to explain that, we have to consider separately the two different scenarios of migration of super-Earths: fast and slow.

When super-Earths migrate slowly (${\rm \tau_{SE} \geq 1 Myr}$) many protoplanetary embryos initially in the inner parts of the disk spiral down onto the star before the super-Earth's passage. Indeed, one could also imagine that planetary embryos scattered by the super-Earths onto distant orbits might type I migrate back into the terrestrial planet region near 1~AU. We find that this process is not very effective. First, very few planetary embryos are scattered on distant orbits when super-Earths migrate slowly (similar to Figures 3 and 7) . Second, in general, those scattered protoplanetary embryos do not have not enough time to migrate from larger distances to the star to the terrestrial planet region because the gas disk dissipates shortly after the migrating super-Earths cross the terrestrial region. Third, even in our extreme cases where  gas lasts relatively longer after super-Earths passage (eg.  ${\rm \tau_{gas}}$=10 Myr and ${\rm \tau_{SE}=3 Myr}$), these migrating planetary embryos  are very few (2 or 3) and rarely are found in the terrestrial planet region at the time the gas is gone. Obviously, as the mass of these bodies are in general $\sim$0.1${\rm M_{\oplus}}$ it is quite unlikely that they be able to collide and accrete  to form an Earth-sized planet near/in the terrestrial zone.

When super-Earths migrate fast, most of protoplanetary embryos and planetesimals survive the super-Earth's passage but planetary embryos continue spiraling toward the star until the gaseous disk dissipates. Thus, during this period (after super-Earths passage until gaseous disk dissipation) many migrating protoplanetary embryos reach heliocentric distance equal to 0.1 AU and are removed from our simulations.  As consequence of this process, these simulations also tend to produce final protoplanetary disks with less amount of mass compared to similar experiments using our fiducial model. However, as advised before, if super-Earths in reality stop at the inner edge of the gas disk many planetary embryos could be saved from the infall onto the star if caught in exterior mean motion resonances with the super-Earths and other embryos. The resonance chain would extend up to $\sim 1$~AU or not depending on the final number of planetary embryos in the chain, and the disk edge location where the migration of the super-Earth stops (or the location of the  farthest super-Earth from the sun in the case of multiplanet system of super-Earths).

\begin{figure}
\includegraphics[scale=.8]{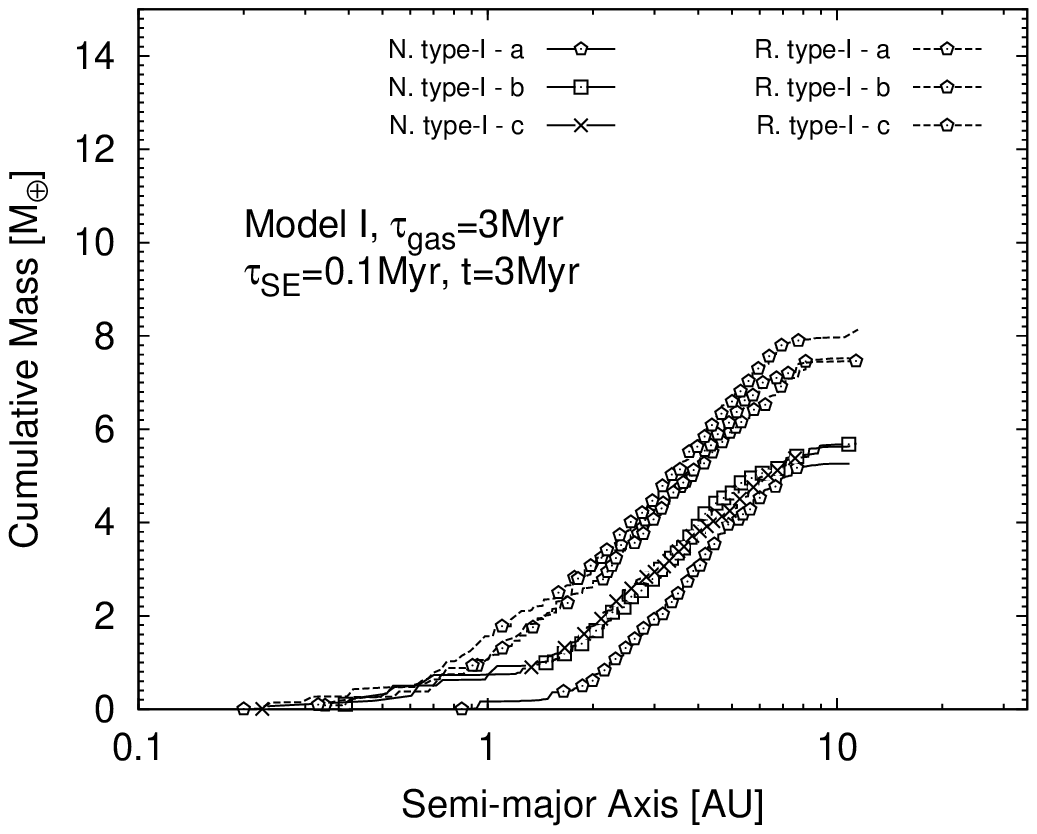}
\includegraphics[scale=.8]{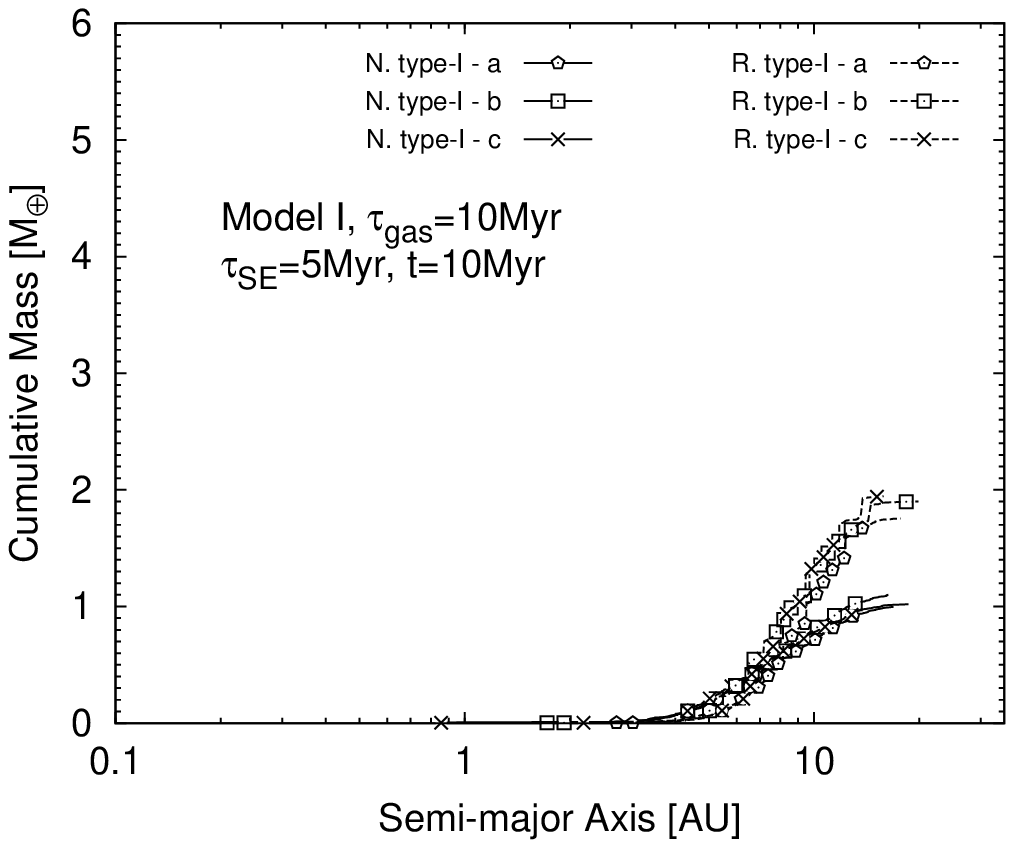}
\caption{Same as Fig. 9 but comparing results obtained imposing different type-I migration timescales on the planetary embryos.  ``N. type-I ' '  labels  represent  simulations using the nominal type-I migration rate for planetary embryos  while the ``R. type-I' ' labels represent simulations where is applied a reduced  type-I migration rate for planetary embryos (see Section 3.1). These results corresponds to simulations within the model I.}
\end{figure}

\subsubsection{Super-Earths Migrating in Sequence}

Instead of migrating in a multiresonant configuration, super-Earths might migrate independently from each other. This phenomenon could be the result of planets forming in sequence at different times. Figure 11 shows snapshots of the dynamical evolution of a system with 6 super-Earths migrating in sequence.  In this case, the migration of super-Earths covers a time span ${\rm \tau_{SE}=5 Myr}$. Each super-Earth migrates inward on a timescale ${\rm \tau_{SE}=5 Myr/6 = 0.83Myr}$. 

\begin{figure*}
\begin{minipage}{\textwidth}
\includegraphics[scale=.63]{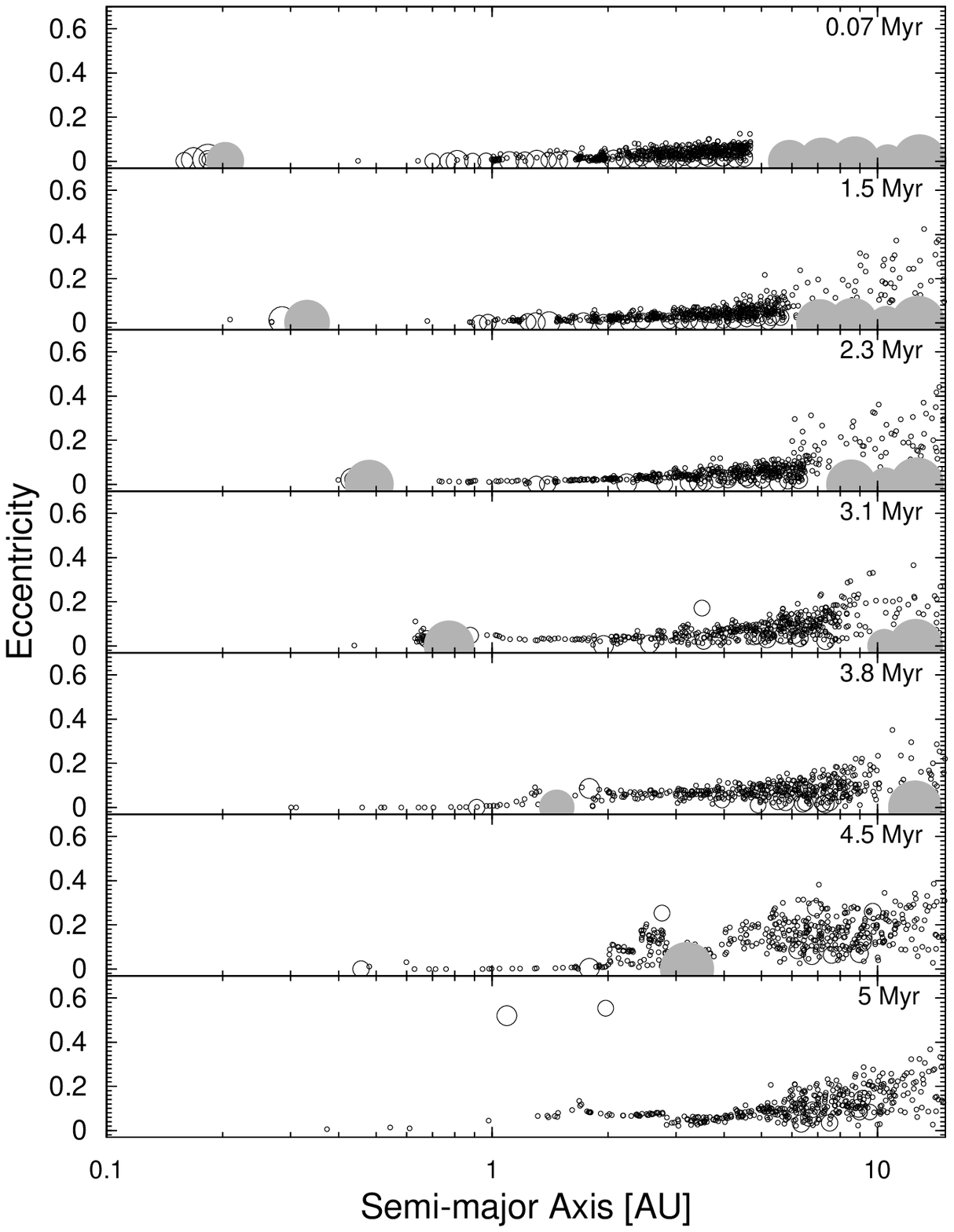}
\includegraphics[scale=.63]{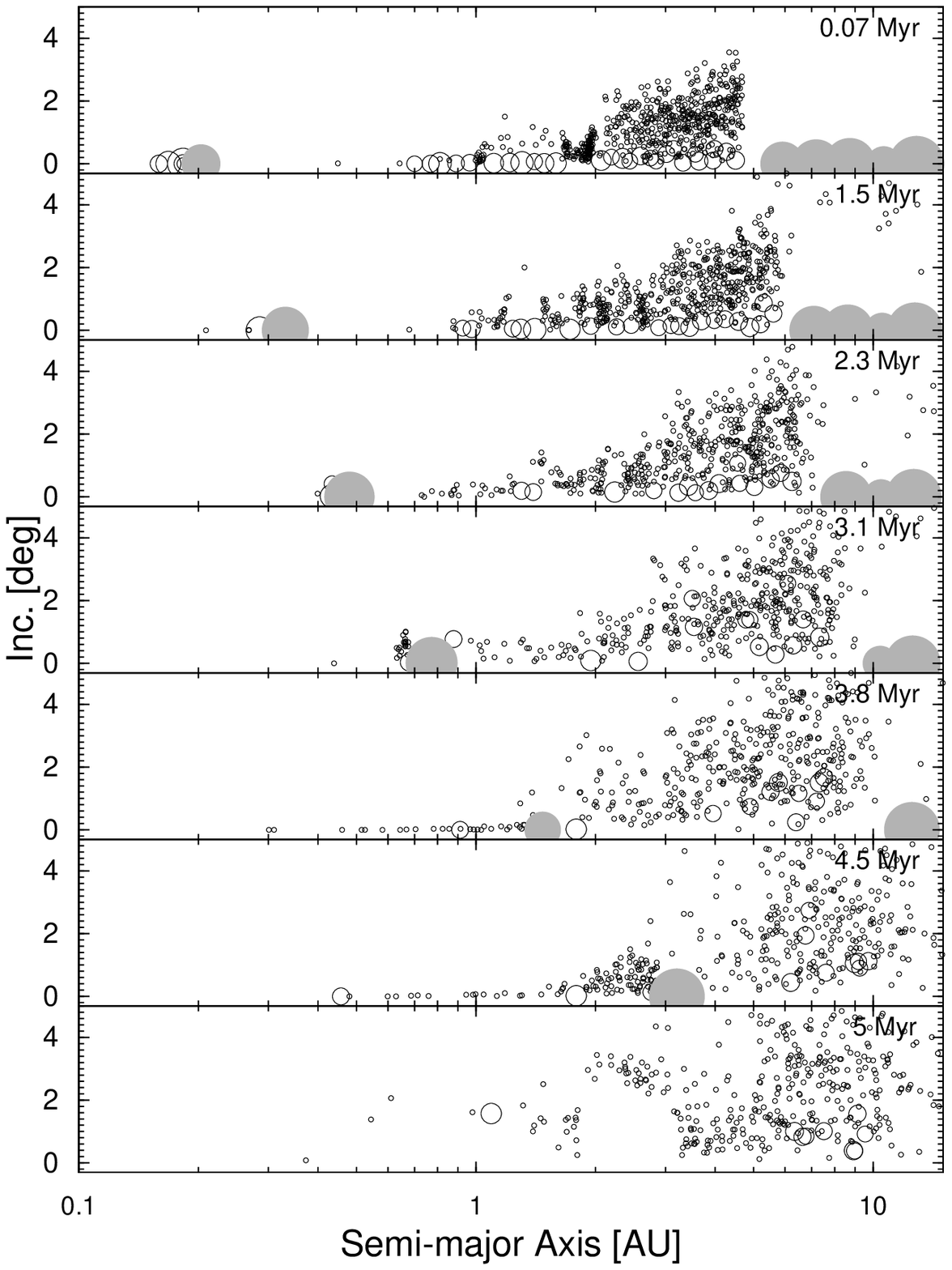}
\caption{Same as Figure 7, but for a system of super-Earths migrating in sequence.}
\end{minipage}
\end{figure*}

When they migrate one by one, each super-Earth triggers its own shepherding and scattering process. After the migration of all super-Earths, only a small fraction of the initial mass in solids survives between 0.5 and 1.5 AU. The effect of the migration timescale is similar for the migration-in-sequence scenario to the locked migration case. If each super-Earth migrates on a timescale smaller than or equal to 0.1 Myr the disk is weakly disturbed. However, if they migrate more slowly then only a small amount of mass tends to survive around 1 AU. Figure 12 shows this using cumulative mass distributions. When the migration timescale is equal to 0.1 or 1 Myr, the results produced by the locked and in-sequence scenarios are indistinguishable. However, for slower migration cases, when ${\rm \tau_{SE}=5}$ or 10 Myr, despite the total surviving mass inside 14 AU being about the same, the simulations with locked migration tend to produce more depletion in the inner parts. This is because the shepherding process is more effective when super-Earths migrate in the locked configuration than in sequence (see Figures 7 and 11). In addition, bodies can be scattered onto more distant orbits if the super-Earths migrate in a locked configuration, because they can be gravitationally passed outward from one super-Earth to the next.

\begin{figure*}j
\begin{minipage}{\textwidth}
\includegraphics[scale=.8]{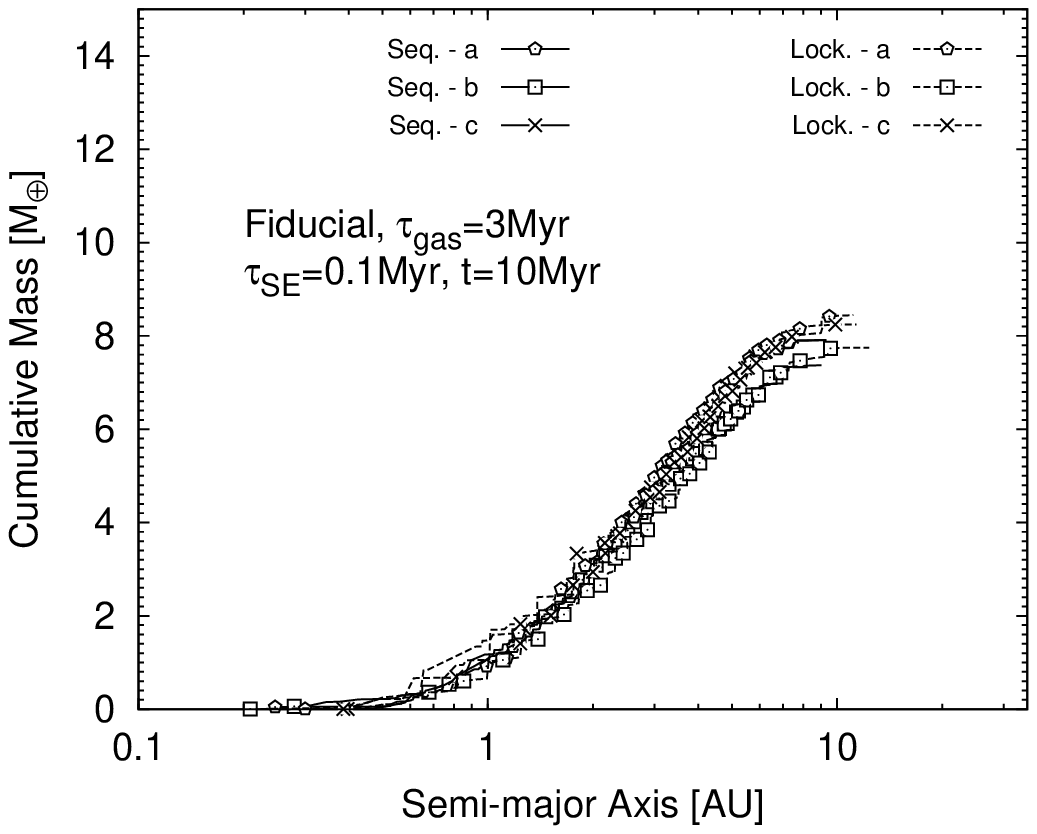}
\includegraphics[scale=.8]{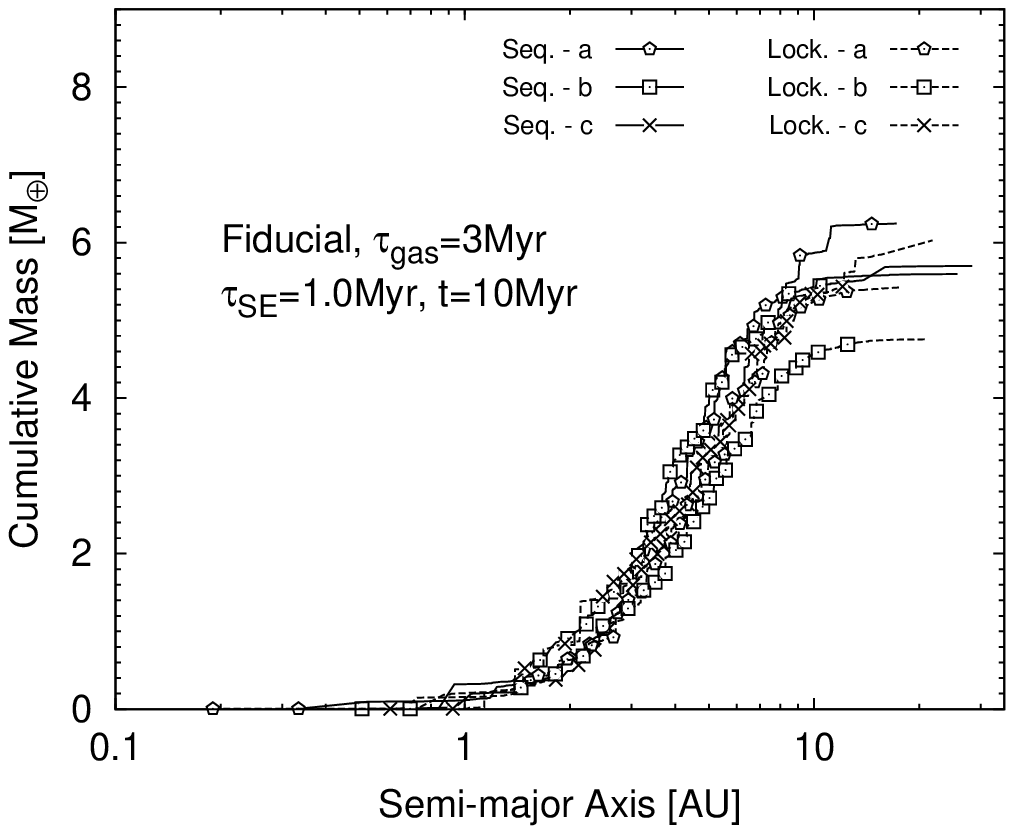}

\includegraphics[scale=.8]{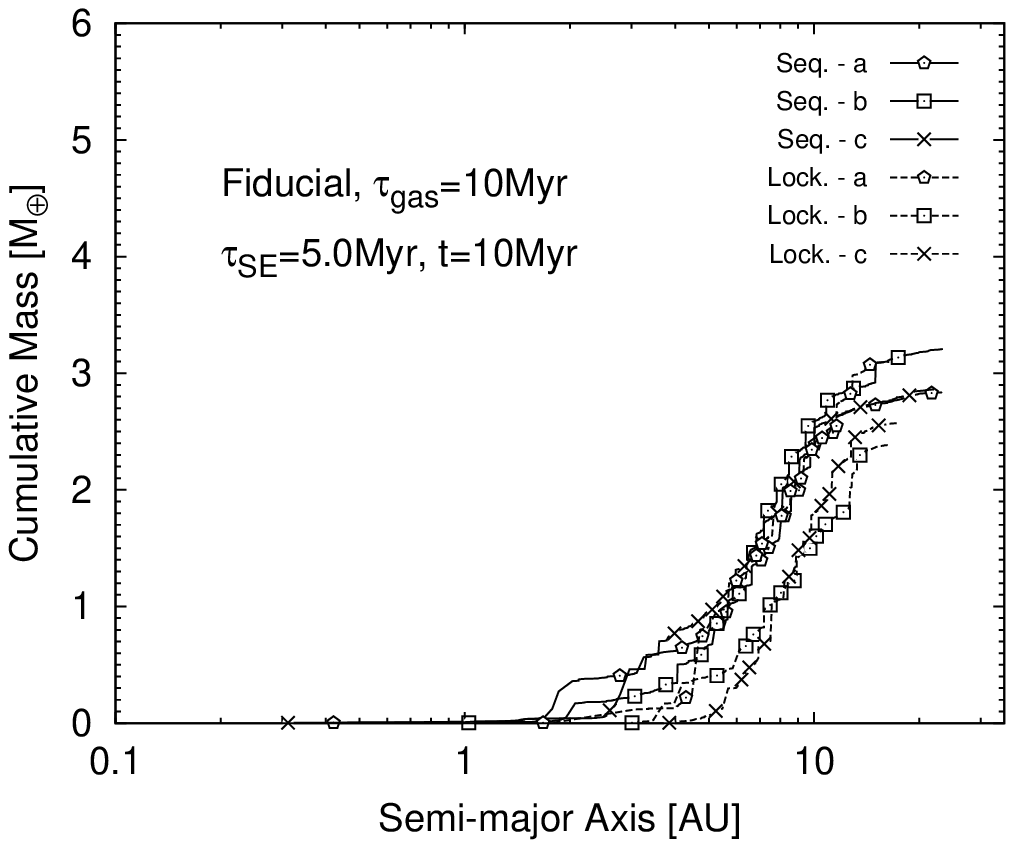}
\includegraphics[scale=.8]{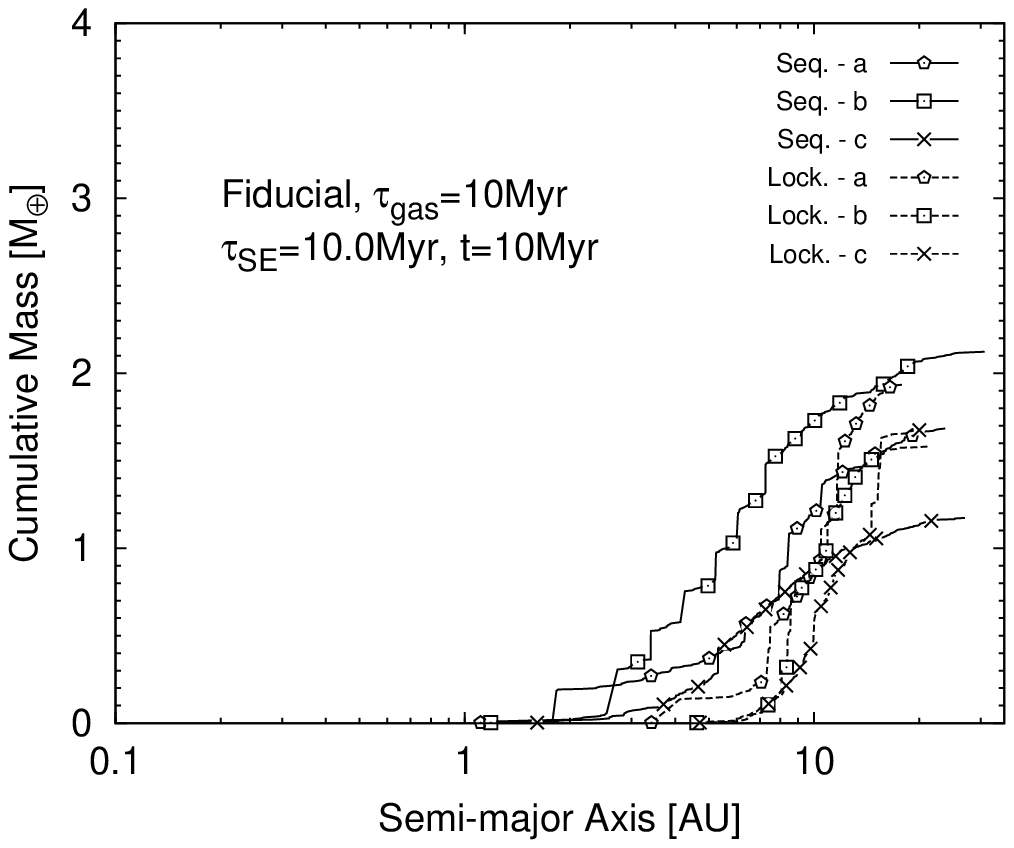}
\caption{Same as Figure 4, but comparing the results of simulations where the super-Earths migrate in sequence versus in a locked configuration. Different panels refer to different migration timescales.}
\end{minipage}
\end{figure*}

\subsection{Radial Mixing}

We now consider the radial mixing of protoplanetary bodies in the disk.  The degree of radial mixing determines each planet's feeding zone, and the feeding zone is the most important factor in determining each planet's composition (e.g., Morbidelli et al 2000; Raymond et al 2007a, b; Izidoro et al 2013; 2014).  

The degree of radial mixing between planetesimals and planetary embryos depends on the super-Earths' migration speed. Figure 13 shows that in simulations where super-Earths migrate fast there is little mixing.  Shepherding of embryos is ineffective when the super-Earths migrate fast. There is some mixing from scattering by sending particles to larger semimajor axis.  In fact, in each bin in Fig. 13 there is little material from farther out but some from close-in. Despite that this configuration of migration does not help to bring volatiles in large quantities from the outer part of the disk to the inner part, the subsequent gravitational interaction between planetesimals and planetary embryos during their final stage of growth may be able to promote some degree of radial mixing of material. Thus, it could still be possible to form Earth-size planets around 1 AU carrying some amount of water and other volatiles delivered by bodies originated from larger heliocentric distances (e.g. Quintana \& Lissauer, 2014).

In contrast, when super-Earths migrate slowly there is significant radial mixing.  This occurs because of the very effective shepherding of material toward the star. Figure 14 shows the radial mixing for the simulations from Figures 3 and 4. The upper panel in Figure 14 correspond to the snapshot in Figure 3 when the time is equal to 4.7 Myr. The lower panel in Figure 14 correspond to the snapshot in Figure 7 when the time is equal to 4.2 Myr.  At these times the planetesimals and protoplanetary embryos have not yet been pushed inside of 0.1 AU and removed from the simulation. As shown in all these figures, most of the planetesimals and planetary embryos initially inward of 4 AU are pushed inside 1 AU  by the super-Earths. Figure 14 shows this compression of mass inside 1 AU.  More than 60-70\% of the total initial mass (${\rm \sim 10M_{\oplus}}$) of the disk carried by protoplanetary embryos and planetesimals is confined inside 1 AU.

Because we are mainly interested in the late accretion of terrestrial-like planets beyond 0.5 AU, in our simulations we remove bodies when they reach orbital radii smaller than 0.1 AU (eg. the last snapshots in Figure 3 and 7). However, in reality, if the super-Earths migration stops (e.g. at the inner edge of the disk), the shepherded material can survive and generate one or more additional planets.  

Migrating super-Earths qualitatively change the compositional gradients in a planetary system.  The super-Earths probably originate beyond the terrestrial planet-forming region and are therefore likely to be more volatile-rich.  When they migrate inward quickly, they pass through the rocky terrestrial planet-forming material and stop at the inner edge of the disk.  Thus, the innermost planets of the system are volatile-rich but the more distant, ``terrestrial'' planets may be rocky.  Even more distant planets are likely to again be volatile-rich.  There is simply a belt that may contain rocky worlds.

The situation is different for slowly migrating super-Earths.  In that case, the material that originated interior to the super-Earths is shepherded inward. In some cases the shepherded embryos may grow massive enough to stop at the inner edge of the disk (Masset et al 2006).  The planets that form from shepherded material have feeding zones that encompass the entire planetary system interior to the innermost super-Earth's starting location.  If super-Earths form at $\sim$5 AU, then this may include a large fraction of material with composition akin to C-type asteroids with significant volatile contents (typically $\sim$10\% water; e.g., Demeo \& Cary 2014).  

Alternately, if shepherded planetary embryos do not grow large enough to halt the migration then they are likely to be destabilized and scattered once they enter the inner parts of the system.  They may be pushed interior to the inner edge of the disk into a gas-free cavity (e.g., Cossou et al 2014) and will likely be accreted onto the super-Earths.  This would slightly ``dilute''  the volatile-rich super-Earths.  Such a system would not contain any rocky planets.  
 
Migrating super-Earths clearly strongly shape the compositional gradient of their parent systems (see also Table 1 of Raymond et al 2008).  Constraints on the compositions of super-Earths around other stars (see Marcy et al 2014) are therefore especially important in understanding the formation of these systems.  The most valuable pieces of information for such studies are compositional constraints on multiple planets in the same system. Once this information is available, the interpretation of each system in terms of formation mechanisms has to be done on a case-by-case basis, but we hope that the analysis presented above will provide some general guidelines to narrow the range of possible scenarios to search.

\begin{figure}
\centering
\includegraphics[scale=.5]{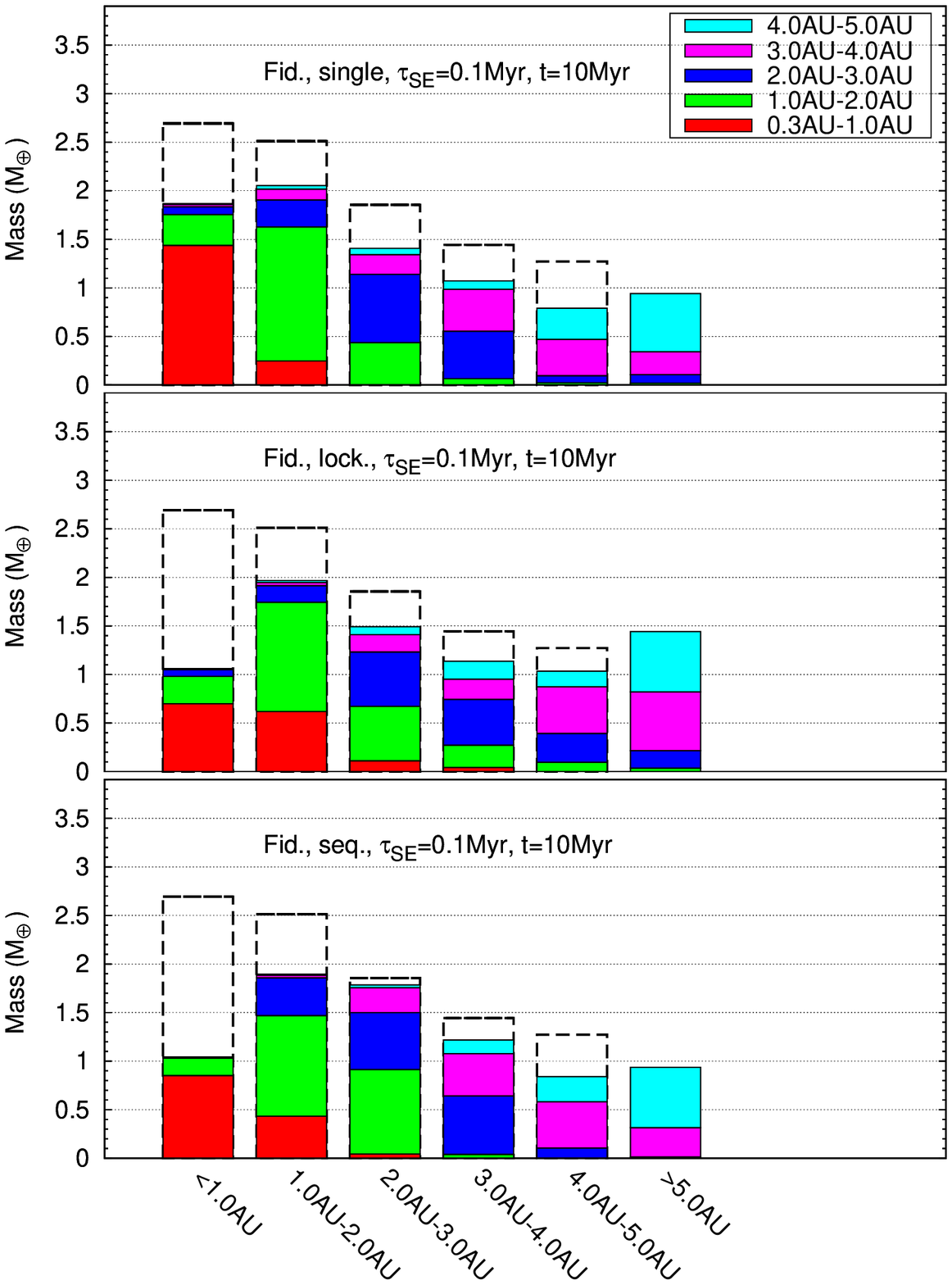}
\caption{Radial mixing of protoplanetary embryos and planetesimals after super-Earth(s) passage for three different configurations of migrating super-Earths. Different configurations are shown on each panel. The super-Earth migration timescale is ${\rm \tau_{SE}}$=0.1Myr. The histogram base shows different regions around the star. Dashed histogram-boxes show the total amount of mass in protoplanetary embryos and planetesimals for such regions in the beginning of our simulations. Each color-histogram-bar is composed by smaller boxes showing the fraction of mass that came from different locations. The colors represent 5 different source regions of material. The sum of the different fractions, for each region shown in the histogram base, gives the respective total amount of mass left after super-Earth passage. These panels show average values computed from 3 simulations  with slightly different initial conditions for protoplanetary embryos and  planetesimals. All these results were obtained within the  fiducial model.}
\end{figure}

\begin{figure}
\centering
\includegraphics[scale=.8]{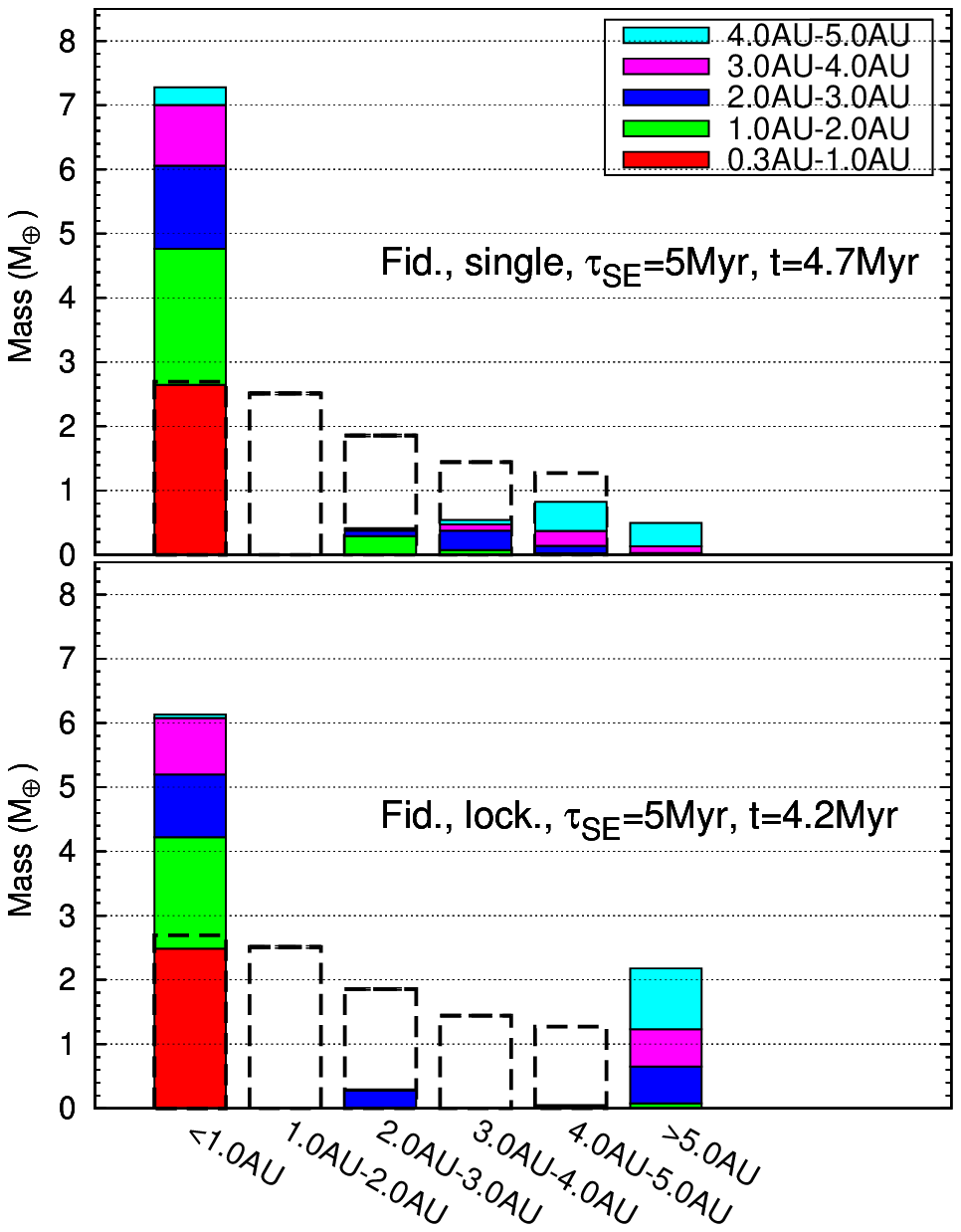}
\caption{Same as Figure 13 but for super-Earths migrating in a timescale ${\rm \tau_{SE}}$=5Myr. Each panel shows the result obtained for a single simulation. The time at which the radial mixing is computed is indicated on each panel and correspond to one of the snapshots shown in Figure 3 (4.7 Myr - upper panel) and Figure 7 (4.2 Myr - lower panel).}
\end{figure}

\section{Discussion}

\subsection{Formation of habitable planets}

Super-Earths (and mini-Neptunes) are known to exist close to at least 30-50\% of main-sequence FGKM stars (e.g., Howard et al 2010, 2012; Mayor et al 2011; Fressin et al 2013).  If these planets formed in situ via the pileup of migrating material (Hansen, 2014; Chatterjee, \& Tan; 2014; Boley \& Ford, 2013) one should expect that terrestrial planets with smaller masses should be present farther out, possibly reaching the habitable zone. However, if super-Earths formed in the outer disk and migrated to their current location,  the effect of their migration must be taken into account when considering the formation of Earth-sized planets in the habitable zone.

The habitable zone of a star (HZ) is defined as the range of heliocentric distances at which an Earth-sized planet could keep liquid water on its surface, given certain assumptions for the planet's atmosphere and mass. A conservative estimate for our solar system suggests that the HZ is between 0.95 and 1.67 AU (Kasting et al., 1993; Selsis et al 2007; Kopparapu et al., 2013, 2014; Leconte et al 2013).  The fraction of Sun-like stars with Earth-sized or super-Earth-sized planets in the HZ has recently been estimated as $\sim 6\%$ (Petigura et al 2013; see also Dong \& Zhu 2013).  This fraction has been estimated as being even higher -- roughly 40\% -- for low-mass stars (Bonfils et al 2013; Fressin et al 2013; Gaidos 2013; Tuomi et al 2014).  

Our results suggest that Earth-sized planets in the habitable zone may not always be Earth-like.  If super-Earths migrated in quickly, then rocky planetary embryos and planetesimals are only modestly disturbed.  Rocky planets like Earth should form readily.  However, if super-Earths migrated slowly, then the rocky material was all pushed into the innermost parts of the system.  Any planets in the HZ are likely to be super-Earth-like and not Earth-like.  In other words, these planets would likely be extremely volatile-rich.  They would be akin to the mini-cores proposed by Raymond et al (2014) rather than true Earth analogues, even at sizes comparable to Earth.  One recent study (Alibert, 2014) suggests that mini-cores are not habitable, because they possess too much water for their mass, so that a layer of high-pressure sub-oceanic ice is expected, isolating the hydrosphere from the crust and preventing a successful CO cycle.  Given the complex feedbacks inherent in any planet (e.g., Kasting \& Catling 2003), we cannot claim to assess the habitability of such non-Earth-like worlds.  We hope that our work can stimulate additional studies in this area.

\subsection{The solar system in context}

With no super-Earths in the inner planetary system and no planets inside of Mercury's orbits, our solar system is an oddity compared to the ``typical'' planetary system.  According to our current understanding of solar system evolution, the peculiar structure of our system is mainly due to three properties:

\begin{itemize}
\item{(i)}~The inner disk was deficient in solid material relative to the outer disk, so that in the inner part only Mars-mass planetary embryos formed before gas removal while the outer part managed to generate at least 4 massive planetary cores.

\item{(ii)}~The innermost two major planets (Jupiter and Saturn) happened to be giant planets with a mass ratio preventing long-lasting inward migration (the so-called Grand-Tack scenario, see Masset and Snellgrove, 2001; Morbidelli \& Crida, 2007; Walsh et al., 2011; Pierens \& Raymond 2011).

\item{(iii)}~The presence of Jupiter and Saturn on orbits not migrating inward prevented Uranus and Neptune from migrating into the inner solar system. Without Jupiter and Saturn, Uranus and Neptune would have become ``typical'' close-in super-Earths (Morbidelli, 2014)
\end{itemize}

It is clear that the properties above, particularly property (II) are nongeneric and require some lucky combination of events.

The specific formation and evolution pattern of the solar system, as improbable as it may be, obviously led to the formation of a habitable and inhabited planet. It is not clear, though, whether this specific pattern is necessary for the emergence of habitable terrestrial planets, or other more generic patterns could be equally good. If the latter were the case, it would be strange that we live in an ``odd'' system, rather than in a typical one.

We have shown that if super-Earths migrate slowly inward then the formation of terrestrial planets is stunted in the habitable zone. In this case, the most typical planetary systems would not host habitable planets; thus something like our solar system is needed, refraining super-Earths from migration. However, if super-Earths form in situ or migrate quickly through the habitable zone this is not the case. Not knowing yet enough about super-Earth formation and migration, unfortunately we cannot conclude on the place of our solar system in the context of habitable planets.

\section{Conclusions}

The {\tt Kepler} mission (Borucki et al 2010; 2011) has discovered a plethora of planets with orbital period shorter than 100 days. Statistical analysis based on the detection rate and bias evaluation indicates that this class of planets should be present in at least 30-50 percent of sun-like stars (e.g., Howard et al 2012; Fressin et al 2013; Petigura et al 2013).  Two models can form this population of planets: in situ accretion and inward migration (see Raymond et al 2008, 2014).  There are a number of reasons to favor the inward-migration model (Terquem \& Papaloizou 2007, Cossou et al 2014).  

In this paper we explore the consequences of the super-Earths migration process for the later accretion of terrestrial planets in these systems.  Type I migration of Earth-mass planets is complicated and sensitive to the disk properties (Paardekooper et al 2011; Kretke \& Lin 2012; Bitsch et al., 2013, 2014).  To cover the range of plausible outcomes we considered different migration speeds for super-Earths.  Fast migration is consistent with isothermal disk models (e.g., Ward 1997; Tanaka et al 2002).  Slow migration assumes that super-Earths are trapped at zero-torque zones that themselves migrate inward on roughly the disk's evolutionary timescale (e.g., Lyra et al 2010, Horn et al 2012).  Simulations were performed considering a single and also a system of migration super-Earths.
 
Fast-migrating super-Earths only weakly disturb the disk of planetesimals and planetary embryos. A significant part of the original mass of the disk survives within its initial semimajor axis distribution. Consequently, the later emergence of terrestrial planets in these systems is very probable. However, in the slow migration scenario, where the migration timescale for the super-Earths is $\sim$1-10 Myr, most protoplanetary embryos and planetesimals are shepherded in resonances by the super-Earths and pushed toward the star. This is particularly the case if the gas effects are strong: tidal damping of the eccentricities and inclinations of the protoplanetary embryos and gas drag on the planetesimals. Of course, given that the super-Earths' migration is gas-driven, gas must be present and shepherding is a natural outcome of slow inward migration of super-Earths.  Shepherding strongly depletes the disk in and near the habitable zone. The subsequent formation of terrestrial-like planets is unlikely, so any planets that form in the habitable zone must be super-Earth-like rather than Earth-like.  The material shepherded inward from interior to the super-Earths' initial orbits ($\lesssim$4 AU) toward the star could stimulate the formation of rocky close-in planets. In that case rocky planets could exist closer in than volatile-rich ones, although neither would have compositions representative of the condensation in their final locations.

\section*{Acknowledgements}

We thank an anonymous referee for the careful reading of our manuscript and the valuable comments. A.~I. thanks financial support from CAPES Foundation (Grant:~18489-12-5). A.~M. and S.~R thank the Agence Nationale pour la Recherche for support via grant ANR-13-BS05-0003 (project MOJO). We thank the CRIMSON team for managing the mesocentre SIGAMM of the OCA, where these simulations were performed.

\end{document}